\documentclass[12pt]{iopart}
\usepackage{graphicx}

\begin{document}
\title{Characterizing spatiotemporal patterns in three-state lattice models}
\author{Matti Peltom\"aki$^1$, Martin Rost$^2$ and Mikko Alava$^1$}
\address{$^1$ Department of Applied Physics, Helsinki University of
  Technology, P.O.~Box 1100, 02015 HUT, Espoo, Finland}
\address{$^2$ Bereich Theoretische Biologie, IZMB, Universit\"at Bonn, 53115
  Bonn, Germany}
\ead{matti.peltomaki@tkk.fi}

\begin{abstract}
A two-species spatially extended system of hosts and parasitoids is studied.
There are two distinct kinds of coexistence; one with populations distributed 
homogeneously in space and another one with spatiotemporal patterns. In the 
latter case, there are noise-sustained oscillations in the population 
densities, whereas in the former one the densities are essentially constants in
time with small fluctuations. We introduce several metrics to characterize the 
patterns and onset thereof. We also build a consistent sequence of corrections 
to the mean-field equations using {\it a posteriori} knowledge from 
simulations. These corrections both lead to better description of the dynamics 
and connect the patterns to it. The analysis is readily applicable to realistic
systems, which we demonstrate by an example using an empirical metapopulation 
landscape.
\end{abstract}


\maketitle

\section{Introduction}

Pattern formation has gained a lot of interest in both the physical
\cite{tainaka94,imbihl95,szabo97,albano98,szabo99,zhdanov99,
zhdanov00,tsekouras02,zhdanov02,szabo02,lele02,provata03,
szolnoki04,traulsen04,
szolnoki05,
reichenbach08}
and the ecological
\cite{solebascompte,gurney98,pascual02,zhang06,nguyen-huu06,mobilia07}
literature. Patterns may either emerge from the dynamics itself or 
reflect the structure of the underlying landscape. Since the mechanism
of pattern formation plays a crucial role, for instance, in ecological 
systems when one is concerned about extinctions of species
\cite{solebascompte,briggs04,reichenbach07},
it is of fundamental interest to be able to pin down the essential 
mechanism. 

The classical model of two-species systems, such as predator--prey or
host--parasitoid systems, is the Lotka--Volterra (LV) model \cite{lotka20}.
It assumes fully stirred, or panmictic, species. However, introducing explicit 
space typically leads to correlated structures where the assumption breaks down
\cite{solebascompte}. The correlations manifest themselves in a wide variety of
forms. The most common ones are spray-like or flame-like in spatial dimensions 
\cite{tainaka94,pascual02,zhang06,mobilia07}, ripple-like in space--time 
\cite{zhang06}, and spiral-like \cite{reichenbach08, gurney98, nguyen-huu06, 
reichenbach07}. Similar patterning has also been observed in several related 
models in statistical physics \cite{szabo97, szabo99, szabo02, tsekouras02,  
provata03, szolnoki05}, in chemical surface catalysis \cite{imbihl95,albano98,
zhdanov99,zhdanov00,lele02,zhdanov02}, in calcium signaling in cells 
\cite{clapham95, berridge97,falcke04}, and in the infamous complex 
Ginzburg--Landau equation (CGLE) \cite{aranson02}, for instance.  

In ecological and chemical systems, the correlations typically weaken the
interactions, since species tend to be aggregated within themselves. They
also provide the prey (host) a spatial refuge since around the prey (host)
there are less predators (parasitoids). This is called the spatial rescue
effect \cite{solebascompte}. Altogether, spatial inhomogeneity can stabilize
the dynamics and strengthen population coexistence \cite{solebascompte,
zhang06,murdoch05} even via several different mechanisms \cite{briggs04}. 
Spatial two-population dynamics can also lead to counter-intuitive effects 
in which increasing the host (prey in prey--predator systems) spreading rate
actually leads to smaller host population sizes \cite{murrel05}. 
Understanding spatiotemporal dynamics in highly fragmented landscapes 
\cite{hanski98} is of even greater difficulty.

There are also myriads of empirical observations of the patterning in
ecological systems. Among the most intriguing examples are field voles
in Northern Britain with evidence of traveling waves \cite{lambin98}, 
mussel beds in the Wadden Sea in the Netherlands with regular spatial patterns
\cite{koppel05} together with voles \cite{ranta97} and lemmings \cite{hanski01}
in Northern Europe. Recently, similar studies have also been performed
in experimental laboratory conditions \cite{koppel08}.

\begin{figure}[!h]
\begin{center}
\includegraphics[width=0.41\textwidth]{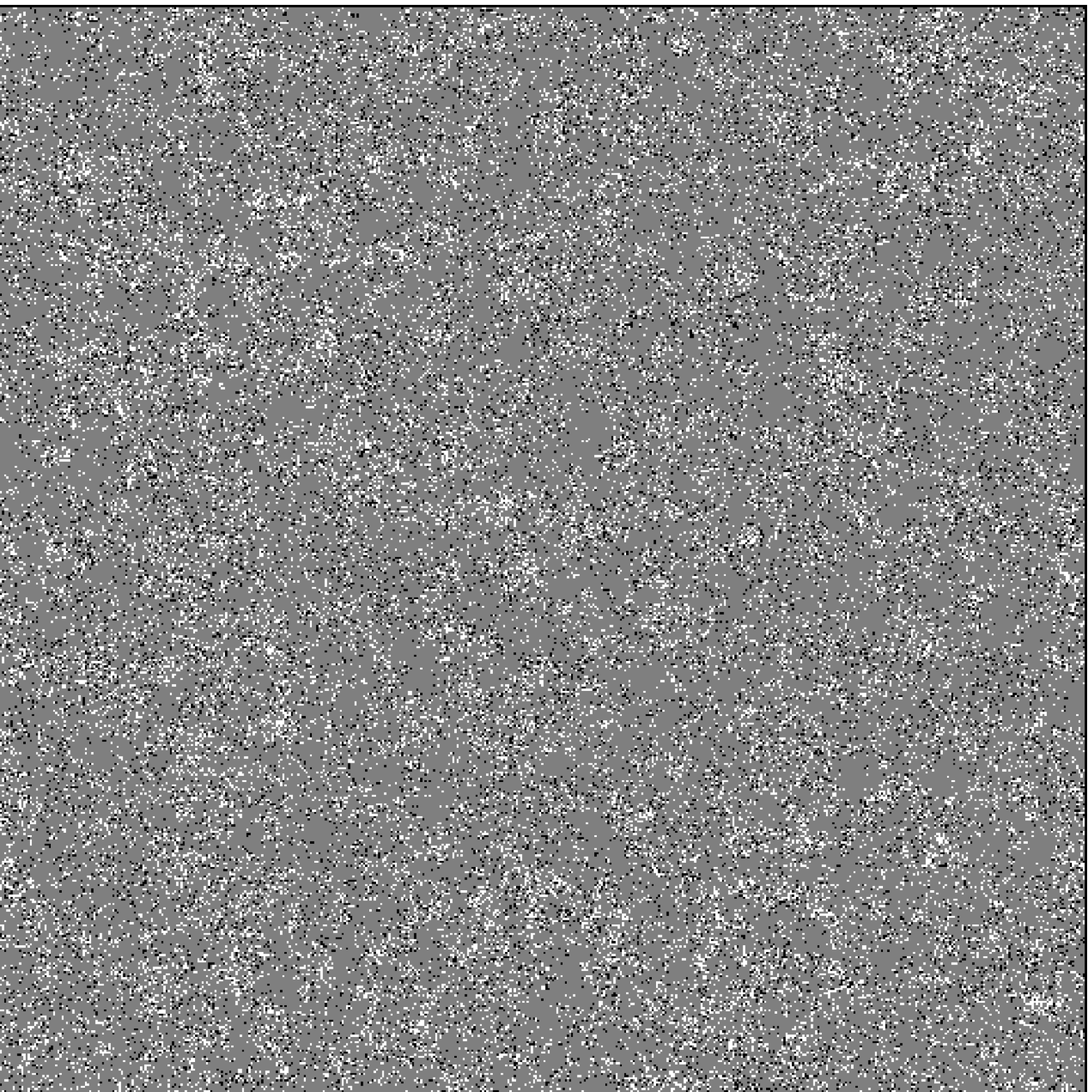}
\includegraphics[width=0.41\textwidth]{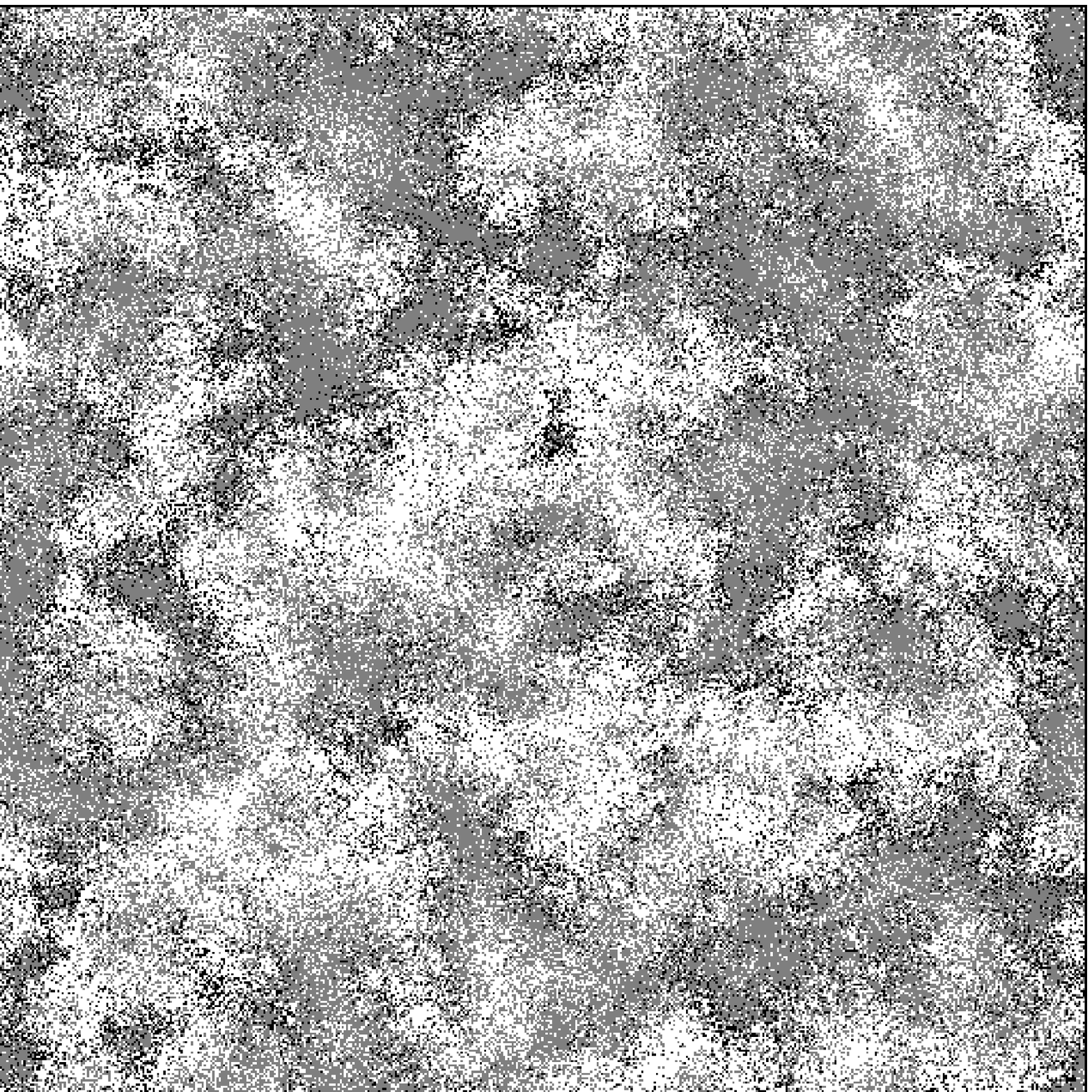}
\end{center}
\caption{Two snapshots of the system with coexistence. Left: disordered 
homogeneous structure for parameters
$w_h=3.0$, $w_p=1.5$, $\lambda_h=0.63$, $\lambda_p=1.3$, and $\delta=0.9$.
Right: a patterned state; parameters are as in the non-patterned one
except for $\lambda_p=2.5$.
Sites with $e$ are white, gray stands for $h$, and
  $p$ is shown in black.}
\label{fig:example}
\end{figure}

In this contribution, we study a numerical model of spatially extended
two-population, or three-state, dynamics. We formulate the model in terms of
hosts and parasitoids but in the scope of this work, these are interchangeable
with prey and predator, respectively. It is defined on a regular square 
lattice, and the spreading of the species occurs in a distance-dependent way 
according to the corresponding incidence functions (see below). There are two 
regimes of coexistence: the species can be either distributed homogeneously 
in space, or build up spatial correlations or patterns 
(see Fig.~\ref{fig:example}).

We study the system from two points of view. The first one concentrates
on the patterns, which we first coarse-grain into areas dominated by one
of the species or empty space. They give rise to vortices as the corner points
of three different areas, and domain walls as the boundary lines between them.
We introduce several quantities describing the geometry and dynamics
of these objects. They include the instantaneous velocities of the vortices, 
their number and lifetime, the average length of the domain walls and 
the shape of the dominance areas close to the vortices. All these quantities
support the division of the parameter space into two kinds of coexistence.

\begin{figure}[!h]
\begin{center}
\includegraphics[width = 8.5cm]{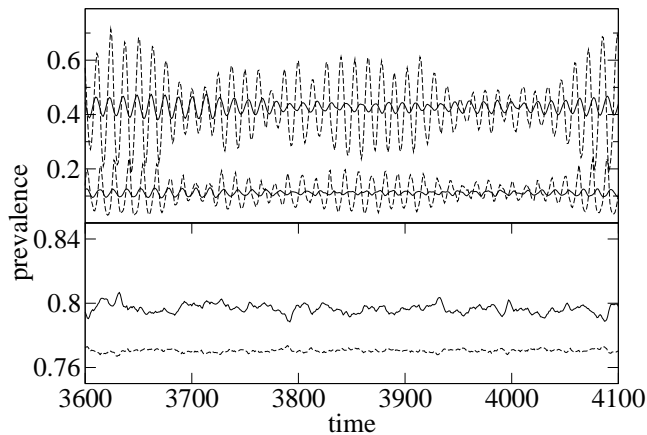}
\end{center}
\caption{Upper panel: the population densities in the patterned state for 
a system of size $L=512$ (solid lines), and a subsystem of size
$L=64$ (dashed lines) of a similar one. The upper curves correspond
to hosts and the lower ones to parasitoids. The parameters are as in
Fig.~\ref{fig:example}.
Lower panel: the population densities in the homogeneous state for the
hosts (solid line) and parasitoids (dashed line) for a system of size
$L=512$. The latter line has been shifted upwards for clarity. The parameters
are as in the upper panel except for $\lambda_p=1.3$.}
\label{fig:example_timeser}
\end{figure}

The second point of view is that of the global dynamics of the system. Here,
we extend on our previous results \cite{peltomaki08} in which we have shown
that for large enough systems there are no non-linearities and in the
spatially correlated coexistence regime, the global dynamics is governed by
noise-sustained oscillations not conforming to the traditional limit cycles. 
See Fig.~\ref{fig:example_timeser} for examples of the corresponding time 
series. We show that the linearization coefficients differ from what one gets
from the complete mixing, or mean-field, assumption, that this is caused by
the dependence of the effective spreading parameters on the instantaneous
global population densities, and that this dependence is vastly different in
the two different coexistence regimes. We also link the dynamics together with
the patterns by showing that the former one can be understood as an infinite
stream of aperiodically occurring spontaneous synchronizations. 

Both the tools for the patterns classification and the analysis of the dynamics
are not restricted to the particular case studied here. Foremost, they are
not restricted to two species (or three states) or two spatial dimensions. In
addition, their domain of applicability does not lie in ecology only; the 
voter models of statistical physics 
\cite{szabo97, szabo99, szabo02, tsekouras02, provata03, szolnoki05} 
and certain surface catalysis reactions 
\cite{imbihl95,albano98,zhdanov99,zhdanov00,lele02, zhdanov02}
lead to similar patterning, and sometimes similar dynamics, as well, and the
machinery discussed here provides means to characterize and classify the
patterns also in these areas. 

This paper is organized as follows. In Sec.~II the model is defined in detail
together with the coarse-graining procedure, vortices, domain walls and
quantities derived from them. The section also discusses the program to
correct the MF equations. In Secs.~III and IV the results regarding patterns
and dynamics, respectively, are presented and discussed. Sec.~V contains an
example application of the analysis, and Sec.~VI discusses the results both 
from the physical and the ecological point of view and makes connections to 
earlier and contemporary literature. Finally, Sec.~VII wraps up the paper and 
concludes. 

\section{Model}

\subsection{Definition}
\label{sec:model_details}

The model describes annual host--parasitoid dynamics on a regular 
two-dimensional square lattice $\Lambda$.
It is inspired by Ref.~\cite{hassell91} but has a wider
interaction range as typical in metapopulation dynamics \cite{hanski98}. 
At a given time, a lattice site can be empty (state $e$), populated by a host
either without ($h$) or with ($p$) parasitoids. The dynamics allows for a 
cycle of transitions, $e \to h \to p \to e \to \dots$. This is a simplification
that neglects the decay of the hosts on their own, and in turn emphasizes the
role of the parasitoids. In particular, spontaneous deaths of non-parasitized
hosts are assumed to be rare enough to be considered nonexistent. Even though 
this means that hosts live forever if parasitoids are absent, the restriction 
is not serious. In coexistence it reduces to assuming faster typical extinction
times for the parasitoids than for the hosts. An equivalent description of the 
model is as a variant of the susceptible-infected-recovered (SIR) model 
\cite{anderson} with longer spreading lengths and rebirth. Here, the infected
state corresponds to hosts, the recovered one to parasitoids, and the rebirth
to the spontaneous death of the parasitoids. 

The transition probabilities between the states are functions of the
{\it connectivities} which in turn depend on the local populations and are
thus directly related to the number of immigrants arriving at a lattice
site. For a given configuration, the connectivity of species $h$ or $p$ on
lattice site $x$ at time $t$ is
\begin{equation} \label{eq:connectivity}
I_{h|p}({\bf x},t) = \sum_{\bf x'} k_{h|p}(|{\bf x}-{\bf x'}|) \;
\chi_{h|p}({\bf x'},t) 
\end{equation}
where $\chi_{h|p}({\bf x},t)$ is the characteristic function, i.e.\ $=
1$ if the state of ${\bf x}$ at time $t$ is $h$ or $p$, respectively, and $= 0$
else. The kernel has an exponential decay with a
characteristic scale $w_{h|p}$ 
\begin{equation}
k_{h|p}(|{\bf x}-{\bf x'}|) \propto \exp \left( -
\frac{|{\bf x}-{\bf x'}|}{w_{h|p}} \right) \, , 
\end{equation}
and is normalized such that its integral over the whole plane equals one. 
In time step $t \to t + 1$, the transition $e \to h$ takes place with
probability $\lambda_h I_h$ and transition $h \to p$ with probability 
$\lambda_p I_p$. The parasitoids die out (the transition $p \to e$)
with probability $\alpha$ irrespective of the surroundings. Parallel updates
are used. There is an absorbing state with the lattice completely filled with
hosts without any parasitoids, together with long-lived reactive coexistence
states.

There are numerous assumptions in the model from the ecological point of 
view. The first and foremost are the long-range interactions. In realistic
cases, they are particularly important for the stability of the system since
they contribute to immigration and survival also in remote habitat patches. 
In a regular lattice, their role is not as emphasized. The assumption, however,
is fairly weak, since the dispersal takes place according to an exponentially
decaying kernel, i.e.~there is a typical short interaction length. Furthermore,
the connectivity (\ref{eq:connectivity}) is defined such that parasitized 
hosts do not contribute to the host connectivity. Naturally, this is not
completely true in empirical systems, since unparasitized hosts can be found
also in parasitized habitat patches and these are fully capable of contributing
to the spreading of the hosts. All these assumptions are minor ones and they
do not hinder one from being able to tackle the relevant and interesting 
properties of the model system.

In the MF approximation, the behavior of the system can be calculated as 
follows. The rate equations for the population densities are
\begin{eqnarray}
h_{t+1} & = & h_t + \kappa(1-h_t-p_t) h_t -
\mu p_t h_t \nonumber \\
p_{t+1} & = & p_t - \delta p_t + \mu p_t h_t \, ,
\label{eq:naivemf}
\end{eqnarray}
where the coupling constants $\kappa$ and $\mu$
are just the total transition rates obtained under constant densities. There
are three fixed points: the coexistence
\begin{equation}
\label{eq:nontrivialfp}
\bar h = \frac{\delta}{\mu} \;\;\; \mbox{and} \; \; \;
\bar p = \frac{\kappa (\mu-\delta)}{\mu(\kappa + \mu)}
\end{equation}
if $\mu \equiv \mu(\bar h, \bar p) > \delta $, the extinction of
parasitoids
\begin{equation}
\bar h = 1 \; \; \; \mbox{and} \; \; \; \bar p = 0
\end{equation}
otherwise, and the extinction of both species
\begin{equation}
\bar h = 0 \; \; \; \mbox{and} \; \; \; \bar p = 0
\end{equation}
which is achieved if the hosts die out before the parasitoids do. The coexistence
fixed point can be either stable or unstable depending on the parameters. The
unstable case corresponds to a limit cycle. Below, we compare these elementary 
observations to the behavior of the spatially extended system. Since the MF 
approximation is merely a rough one, there is no particular reason to expect
one-to-one correspondence.

\subsection{Characterizing patterns}
\label{sec:characterizing_patterns}

\begin{figure}[!h]
\begin{center}
\includegraphics[width = 4cm]{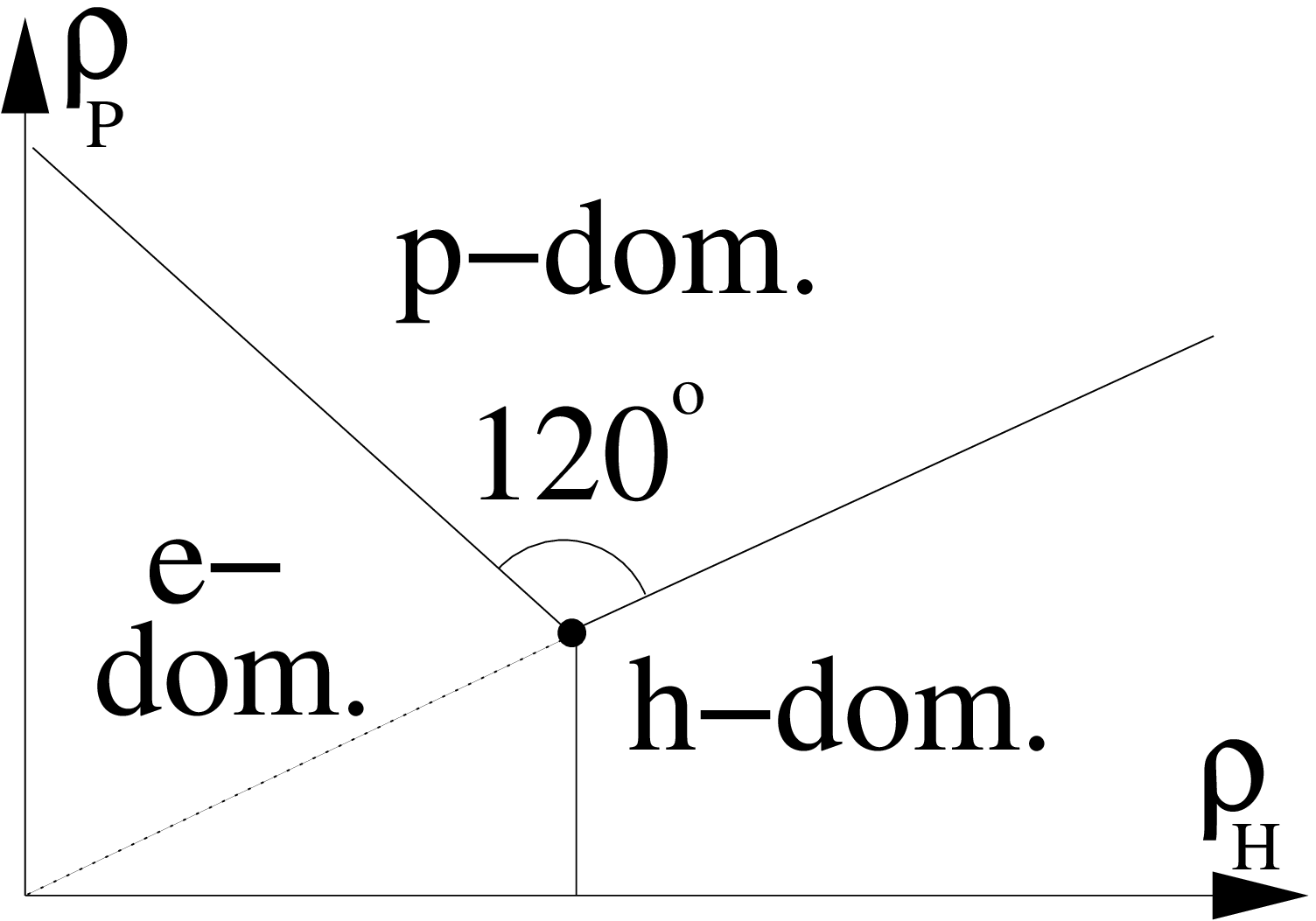}
\hspace{1cm}
\includegraphics[width = 4cm]{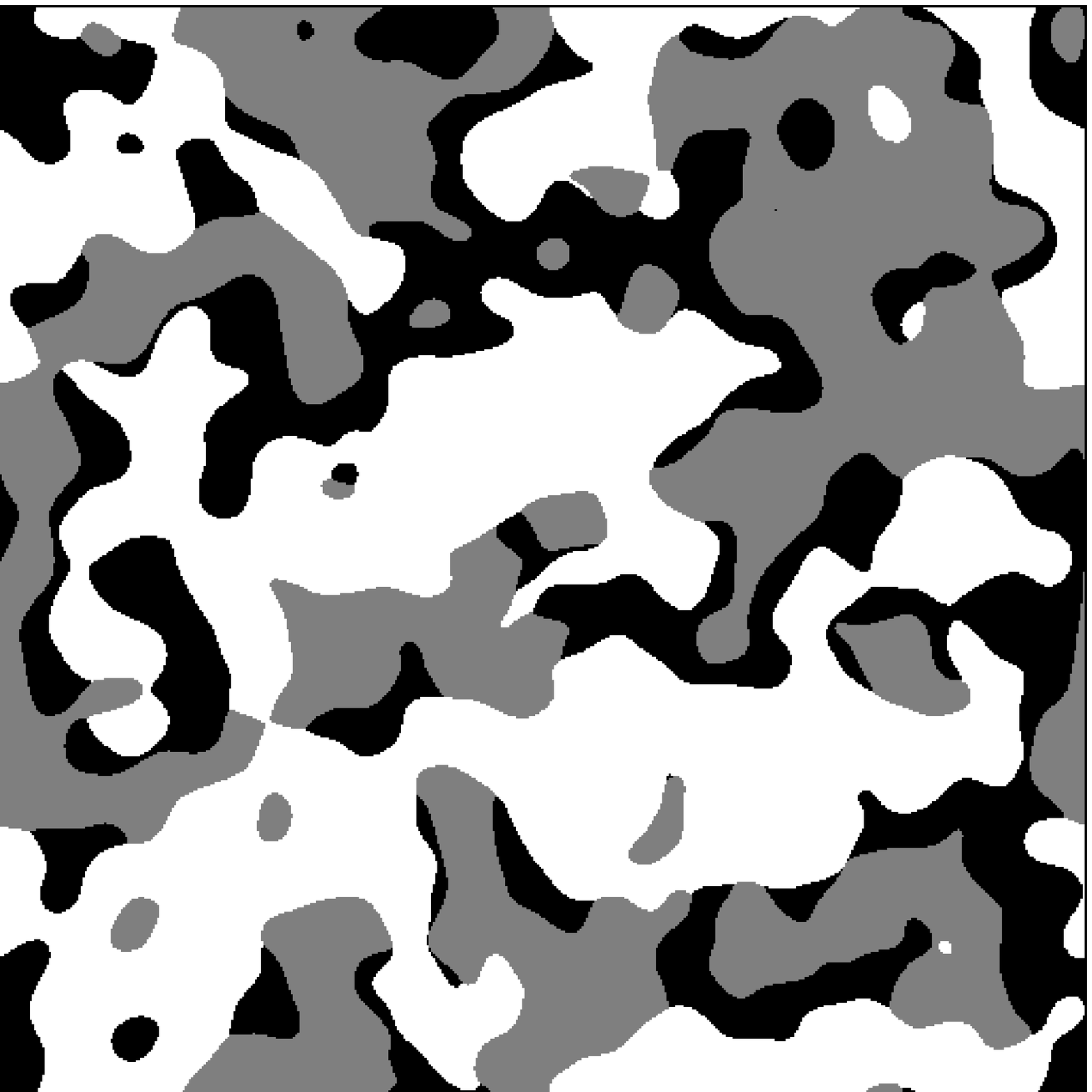}
\end{center}
\caption{
(a) The division of the phase space into three sectors associated with the 
three states, $e$, $h$ and $p$. For given average densities
  $\bar h$ and $\bar p$, the first quadrant of the 
  $(\rho_h,\rho_p)$-plane
  is divided into three regions by three lines. That separating $h$- and 
  $p$-dominated regions starts from the average $(\bar h, \bar p)$ (the
  black dot), goes towards increasing $\rho_h$ and
  $\rho_p$ and is such that its continuation (the dotted line)
  passes through the origin. The other two lines form 120-degree angles with
  the first one and each other. (b) The computed dominance regions of the
  configuration shown in Fig.~\ref{fig:example}.}
\label{fig:domains}
\end{figure}

To characterize the patterns, consider a spatially smoothed continuous
field of population densities \cite{peltomaki08}
\begin{equation}
\rho_{h|p,w}({\bf x},t) \equiv \sum_{x'} k_w(|{\bf x}-{\bf
  x'}|) \; \chi_{h|p} ({\bf x'},t)  \, ,
\label{eq:smoothed_densities}
\end{equation}
where the smoothing kernel $k_w({\bf x})$ is 
a two-dimensional Gaussian such that
$\sum_{\bf x} k_w({\bf x})  = 1$, 
$\sum_{\bf x} {\bf x} k_w({\bf x}) = {\bf 0}$, and 
$\sum_{\bf x} {\bf x}^2 k_w({\bf x}) = w^2$. Here the variance $w$ is called 
the smoothing width, and the summation runs over all lattice sites.
At each site ${\bf x}$, the smoothed densities $\rho_{h|p,w}({\bf x},t)$ 
oscillate around the temporally and spatially averaged densities 
\begin{equation} \label{eq:density_average_host}
\bar h = \frac{1}{N}\lim_{T \to \infty} \frac{1}{T}
\sum_{t=1}^T \sum_{\bf x} \chi_h({\bf x},t)
\end{equation}
and
\begin{equation} \label{eq:density_average_parasite}
\bar p = \frac{1}{N}\lim_{T \to \infty} \frac{1}{T}
\sum_{t=1}^T \sum_{\bf x} \chi_p({\bf x},t) \, .
\end{equation}
We use these to divide the two-dimensional phase space
spanned by the densities into three sectors, defined in the caption of 
Fig.~\ref{fig:domains}. They 
are then used to divide the system into {\it prevalence domains} so that
site ${\bf x}$ at time $t$  is defined to belong to a domain according to
the region of the phase space containing the smoothed densities at 
location ${\bf x}$, i.e.~$(\rho_h({\bf x},t), \rho_p({\bf x},t))$ in 
Eq.~(\ref{eq:smoothed_densities}). The regions coarse-grain on a scale 
somewhat larger than the typical interaction lengths. They are not sensitive 
to changes in the smoothing width $w$ given that it is larger than the 
interaction lengths and smaller than the system size $L$.

The domains readily give rise to more sophisticated quantities. To start,
define the vortices as the corner points of the three different types of
domains. These are associated with a positive or negative unit ``charge''
since one can encounter the three domains in two different orders by 
traversing a circle around it in a given direction (say, counter-clockwise). 
Similar structures with three kinds of domains rotating around their corner
points without any coarse-graining have been identified earlier in statistical
physics in the context of the three-state voter model \cite{szabo99}, 
the extended three-state voter model and the three-state Potts model
\cite{szolnoki04}, and combinations thereof \cite{szabo02,szolnoki05}. Also
a four-state model with game-theoretical inspirations \cite{traulsen04} 
shows similar vortices. In three 
dimensions, the vortices generalize to strings \cite{tainaka94}. Further, the 
domain walls are defined as the boundary lines between two domains of 
different types. There are three kinds of walls, according to the possible 
three pairs of domain types they separate, and exactly one wall of any given 
type emanates from a given vortex. The vortices and domain walls provide 
simple means to describe the population patterns, as we demonstrate below.

To derive quantities that follow the evolution of the vortices and domain 
walls in time, it is necessary to be able to reliably track their motion. 
To this end, we have implemented the following procedure. 
Let the $A_-$ and $A_+$ be the sets of 
vortices with negative and positive sign, respectively, at time $t$. Denote
by $B_-$ and $B_+$ the same after one discrete time step, i.e.~at time
$t+1$. An interpretation for the movements of the vortices, including
creation and annihilation or pairs, is a simple pairing of the sets
$X = A_- \cup B_+$ and $Y = A_+ \cup B_-$. If an element $(x,y)$ of this 
pairing consists of vortices in $A_-$ and $B_-$ ($A_+$ and $B_+$) a 
negatively (positively) charged vortex is interpreted to have moved but not
annihilated. If the element consists of vortices in $A_-$ and $A_+$, the two
vortices are considered to have annihilated with each other, and, finally, 
if the element is a pair of vortices from $B_-$ and $B_+$, the vortices are
considered to be newly created vortices at time step $t+1$ that did not 
exist at time $t$.

There is an enormous number of such pairings, and physically motivated means
of selecting one are needed. To this end, define a fully connected weighted
bipartite graph $G=(X,Y,E,w)$, where the vortex sets 
$X$ and $Y$ are the vertices, the edge set their Cartesian product 
$E = X \times Y$, and the weight
function $w$ assigns each edge ($x,y$) a weight equal to the Euclidean 
distance between vortices $x$ and $y$. We use the so-called
Hungarian method from graph theory to obtain the pairing of the sets $X$ and 
$Y$ that has the minimum total weight. This pairing is then used to interpret 
the motion, annihilation, and creation of the vortices as described above. The 
Hungarian method itself is a well-known exact optimization algorithm for the 
bipartite pairing problem based on graph flows. It is fairly complex but it 
runs in cubic time with respect to the number of elements in the sets to be 
paired. Detailed explanations of it can be found in the literature 
\cite{jungnickel,ahuja}.

Building on the tracking procedure for the vortices, it is simple to 
outline a similar procedure for the domain walls. We consider a given
wall observed at time $t$ to be the same wall as another one at time 
$t+1$ if and only if it is of the same type (separates the same two
domain types) and if both its endpoint vortices are identified as the
same according to the method above. A necessary condition for
this is that the vortices have not been annihilated.

\begin{figure}[!h]
\begin{center}
\includegraphics[width = 5cm]{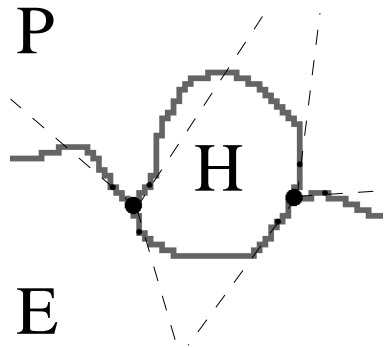}
\end{center}
\caption{An illustration of the vortices, the domain walls, and the 
tangents of them describing the geometry of the walls near the vortices. 
The background is 
a magnified portion of Fig.~\ref{fig:domains} with all three domains
colored white and marked with capital letters, and the boundaries marked
gray. The locations of the vortices are shown with thick black circles. The
smaller black circles show lattice sites on the wall that are at the distance 
of $w_t = 5$ lattice units from the vortices, and the dashed lines drawn
via the vortices and these points are the domain wall tangents.}
\label{fig:schematic}
\end{figure}

After implementing these identification procedures, the vortices and walls
can be used to derive a practically unlimited number of quantities that,
in a way or another, describe the dynamics or the statics of the patterns. 
Here, we have used several, selecting the ones considered most suitable. 
These include the jump length of the vortices, i.e.~distance traveled 
by a vortex in unit time, the number and lifetime of vortices, the geometry
of the domains in the immediate vicinity of the vortices described by the 
widths of the sectors the domains form around them, and the lengths of the
domain walls themselves. See Fig.~\ref{fig:schematic} for a schematic 
illustration of some of these quantities.

Upon applying these metrics, attention must be paid to the fact that random
fluctuations around the average densities also create vortices and domain 
walls. This is an unfortunate side effect of the definitions that must not be 
neglected but has to be taken into account. This is especially important in the
case of the domain wall lengths, in which the fluctuations give rise to a 
nonzero background level, and it is the deviations of the wall length 
statistics from this background -- not from zero -- that signals the onset of 
patterns. Here, we take the background into account by creating random 
configurations of the system by assigning each lattice site a state of $e$, 
$h$, or $p$ randomly using the spatiotemporally averaged densities $\bar h$ and
$\bar p$ from a simulation as the occurrence probabilities of the respective 
states. The background vortex and wall length levels are then obtained by 
computing the same metrics for these random configurations.

\subsection{Effective mean--field dynamics}
\label{sec:effmf}

The MF equations (\ref{eq:naivemf}) discussed in the Introduction assume full 
mixing both within a single species and between two species. In spatially 
extended systems, such as the one studied here, the assumptions fail almost 
invariably. There are several known ways to incorporate spatial effects. These 
include, for example, pair approximations \cite{ben-avraham92,satulovsky94,
peltomaki05,tome07,arashiro08}, rescaling the coordination number of the 
lattice \cite{pascual01}, and others \cite{ovaskainen06}. Here, our approach is
to use input from numerical simulations to introduce corrections of different 
order to the MF equations. These corrections are such that there is no need
to specify a functional form for the interactions before the analysis. 

The zeroth-order correction is to consider the change in the effective
values of the spreading rate parameters introduced by the spatial effects.
This is done by running the simulations for different values of the
parameters, obtaining the average population densities as a function of them,
and subsequently using the inverse of Eqs.~(\ref{eq:nontrivialfp}) to arrive
at those effective values of the MF parameters that lead to the densities
observed in the simulations. However, common experience suggests that 
this kind of rescaling of the parameters is an utterly implausible explanation
for the spatial effects. Instead, higher order schemes have to be used. 

The first-order correction builds on the zeroth-order one. Now, in addition,
the effective parameters are considered functions of the instantaneous
population densities $h_t$ and $p_t$. 
For $\lambda_\alpha I_\alpha({\bf x},t)$ small, they equal the spreading
probabilities, and thus dynamics can be written as
\cite{peltomaki08}
\begin{equation}
h_{t+1} = \! h_t + \! \! \sum_{{\bf x} \in \Lambda} \! \Bigl[
  \lambda_h k_h({\bf x}) C_{eh}({\bf x},t) - \lambda_p k_p({\bf x})
  C_{hp}({\bf x},t) \Bigr] \nonumber
\end{equation}
and
\begin{equation}
p_{t+1} = (1\!-\!\delta) p_t + \lambda_p
  \sum_{{\bf x} \in \Lambda} k_p({\bf x}) \; C_{hp}({\bf x},t) \, ,
\label{eq:LV}
\end{equation}
where the influence of the connectivities on the prevalence dynamics is 
expressed by the correlation functions
\begin{equation}
C_{\alpha \beta}({\bf x},t) = \frac{1}{|\Lambda|} \sum_{{\bf x'} \in
  \Lambda} \chi_\alpha({\bf x'},t) \; \chi_\beta({\bf x}+{\bf x'},t) \, .
\end{equation}
An approximation of these equations can be written as
(cf.~Eqs.~(\ref{eq:naivemf}))
\begin{eqnarray}
h_{t+1} & = & h_t + \kappa(h_t,p_t) \left(1-h_t-p_t\right) h_t -
\mu(h_t,p_t) p_t h_t \nonumber \\
p_{t+1} & = & p_t - \delta p_t + \mu(h_t,p_t) p_t h_t \, .
\label{eq:generalmf}
\end{eqnarray}
This is an approximation of the usual MF form with the interaction parameters
$\kappa(h,p)$ and $\mu(h,p)$ generalized to be arbitrary functions of the 
instantaneous densities. The full form of them seems to be in general not 
obtainable analytically. In general, they can be nonlinear, and in particular 
they do not have to conform to the standard LV model nor to any ad-hoc 
approximations discussed in the literature, for example turning the 
coordination number of the lattice into two effective coordination numbers that
serve as phenomenological parameters separately for the hosts and the 
parasitoids \cite{pascual01}. To arrive at the first-order correction, perform 
a series expansion of Eqs.~(\ref{eq:LV}) around the fixed point 
$(\bar h, \bar p)$ and introduce auxiliary variables $\eta_t$ and $\pi_t$ such 
that $h_t = \bar h + \eta_t$ and $p_t = \bar p + \pi_t$. Doing this, we arrive 
to linear order at
\begin{equation}
\left(
\begin{array}{c} \eta_{t+1} \\ \pi_{t+1} \end{array} 
\right)
=
\left(
\begin{array}{cc}
a_{h,h} & a_{h,p} \\
a_{p,h} & a_{p,p}
\end{array}
\right)
\left(
\begin{array}{c} \eta_t \\ pi_t \end{array}
\right)
\label{eq:lin}
\end{equation}
with the matrix elements
\begin{eqnarray}
a_{h,h} & = & 1 \! + \! \kappa \! - \! 2 \kappa \bar h \! - \! (\kappa
  \! + \! \mu) \bar p \! + \! \partial_h \kappa \, \bar h (1 \! - \!\bar
  h\!-\!\bar p) - \partial_h \mu \, \bar h  \bar p \nonumber \\
a_{h,p} & = & -(\kappa \! + \! \mu) \bar h - \partial_p \kappa \; \bar
  h (1\!-\!\bar h\!-\!\bar p)  - \partial_p \mu  \; \bar h  \bar p
  \nonumber \\
a_{p,h} & = & \mu \bar p + \partial_h \mu \; \bar h \bar p 
\label{eq:matrix_elements}
\\
a_{p,p} & = & 1 - \delta + \mu \bar h + \partial_p \mu \; \bar h \bar
  p \nonumber \, ,
\end{eqnarray}
in which $\mu$, $\kappa$, their derivatives, and the population densities
are evaluated at the fixed point $(\bar h, \bar p)$.

If the four derivatives are set to zero in Eq.~(\ref{eq:matrix_elements}),
Eqs.~(\ref{eq:lin}) and (\ref{eq:matrix_elements}) fall back to the
mean-field approximation, and the matrix in Eq.~(\ref{eq:lin}) is directly 
the linearization matrix usable for the standard stability analysis. On the
other hand, if the derivatives are nonzero, the mean-field analysis is not
valid anymore. This situation can be interpreted as instantaneous densities
affecting the spreading rates. 

As we have demonstrated earlier \cite{peltomaki08} and will elaborate on below,
the linear form of Eqs.~(\ref{eq:lin}) and (\ref{eq:matrix_elements}) can be 
readily fitted to temporal data from simulations of the spatially extended 
model, and numerical values for the matrix elements can be easily obtained. 
This allows one to solve for the four derivatives, $\kappa$ and $\mu$ with 
respect to $h$ and $p$, and subsequently arrive at the first order correction 
to the MF equations. Note that this computation reveals the domain of 
applicability of the mean-field treatment by either yielding vanishing 
derivatives (applicable) or non-vanishing ones (not applicable). In principle, 
the procedure could be continued \emph{ad infinitum} to higher derivatives 
leading to higher-order corrections. These calculations are omitted here, 
however, since the linear treatment turns out to be enough to demonstrate and 
discuss the effect of the corrections. Note that the program outlined here is 
by no means restricted to the present model, but it can be applied to virtually
any other one with a similar structure. In particular, it is not restricted to 
two-species (or three-state) models, two spatial dimensions and not even to 
discrete time, since continuous-time results can always be retrospectively 
discretized by sampling snapshots, nor discrete space. Promising candidates for
such an application include those in Refs.~\cite{szabo97,szabo99,szabo02, 
tsekouras02,
provata03,szolnoki05,antal01a,antal01b,wood06prl,wood06}, for 
example.

\section{Patterns}
\label{sec:lines}

\begin{figure}[!h]
\begin{center}
\includegraphics[width = 7cm]{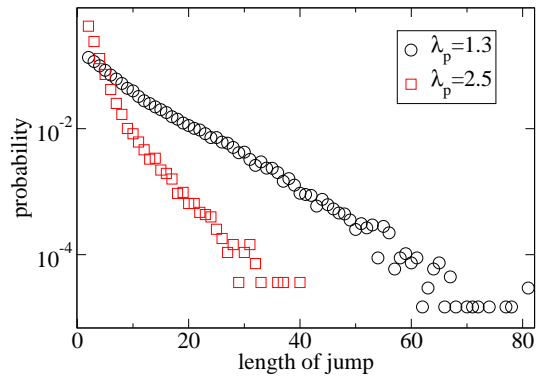}
\end{center}
\caption{Distribution of the jump length of the vortices, 
i.e.~the distance
traveled by a vortex in unit time, in both the non-patterned
($\lambda_p=1.3$) and the patterned ($\lambda_p=2.5$) case. Other parameters
are as in Fig.~\ref{fig:example}. The average jump lengths are approximately
3.6 and 9.2 lattice units for $\lambda_p=2.5$ and $\lambda_p=1.3$,
respectively.}
\label{fig:vortex_jump_length}
\end{figure}

\begin{figure}[!h]
\begin{center}
\includegraphics[width = 7cm]{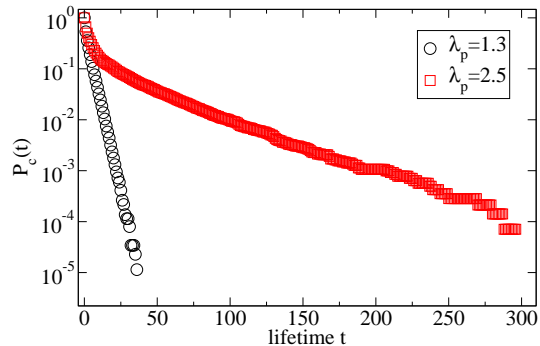}
\end{center}
\caption{Cumulative distribution of the vortex lifetime according to the 
tracking algorithm in non-patterned and patterned cases. See the caption of 
Fig.~\ref{fig:vortex_jump_length} for the values of the parameters. The 
average lifetimes are approximately 2.9 and 8.7 discrete time units for 
$\lambda_p=1.3$ and $\lambda_p=2.5$, respectively.}
\label{fig:vortex_lifetime}
\end{figure}

The vortices and the domain walls can be located and followed through time 
as described in Section \ref{sec:characterizing_patterns}. In general, 
considering quantities derived from these leads to an observation of two 
distinct kinds of coexistence regimes. There is one with formation of chaotic 
moving fronts, patterns, and another one without them. All computations 
discussed in detail below support the division.

First, we consider the jump length of the vortices, i.e.~the distance a
given vortex moves in unit time. A histogram of these is plotted in
Fig.~\ref{fig:vortex_jump_length} for the values of the parasitoid spreading
rate parameter $\lambda_p$ corresponding to the patterned and the 
non-patterned cases. Both distributions are exponential, but the patterned
case has a clearly smaller average jump length than the non-patterned one.
Similarly, we plot the lifetime distributions of the vortices in the same
two cases in Fig.~\ref{fig:vortex_lifetime}. Now, the difference between
them is even bigger; in the patterned case the vortices live typically ten
times as long as in the non-patterned case. Therefore, with patterns the
vortices are more stable than without them both with respect to movements
and annihilation. Were the typical vortex jump lengths considered in units
of the typical pattern size (for example, the average domain wall length
discussed below), the difference between the two cases would be even more
pronounced.

In a more detailed inspection, the vortex lifetime distribution in the 
patterned case ($\lambda_p=2.5$) in Fig.~\ref{fig:vortex_lifetime} is composed 
of two parts; for short lifetimes the distribution follows that of the 
non-patterned case, and for long lifetimes, there is a broad tail. A direct 
consideration of the definition of a vortex reveals the cause. Namely, vortices
can, roughly speaking, originate in two different ways. They can be either 
signs of a ``real'' inhomogeneous structure such as seen in 
Fig.~\ref{fig:example}, or merely random fluctuations around the average 
population densities. The former leads to the wide tail, whereas the latter
shows up as the short-lifetime part of the distribution.

\begin{figure}[!h]
\begin{center}
\includegraphics[width = 7cm]{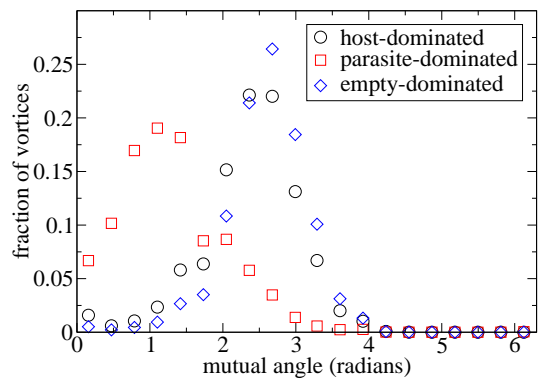}
\end{center}
\caption{The distributions of the angles between the tangent 
directions
of the three domain walls at a vortex (see Fig.~\ref{fig:schematic}).
The angles are multiplied by the vortex
charge so that for each vortex the three domains are passed in the same
direction. The data is from the patterned state, and parameters are as in
Fig.~\ref{fig:example}.} 
\label{fig:mutual}
\end{figure}

\begin{figure}[!h]
\begin{center}
\includegraphics[width = 7cm]{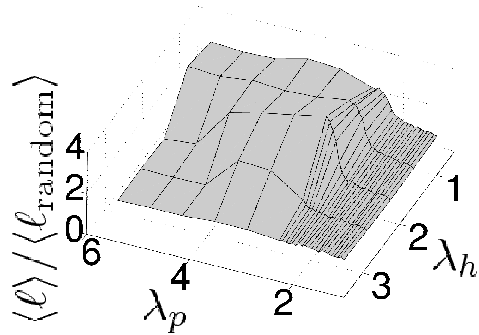}
\end{center}
\caption{The ratio of the measured average domain wall length to its
counterpart in random configurations as a function of the parasitoid and host
spreading rate parameters $\lambda_p$ and $\lambda_h$. The ratio is
approximately one in the uniformly distributed case and increases to 
approximately three in the patterned case. In all cases, the distribution
of the domain wall lengths decays exponentially (not shown).
Other parameters are as in Fig.~\ref{fig:example}.}
\label{fig:ll}
\end{figure}

Next, we turn to the geometry of the dominance regions in the vicinity of 
the vortices. We compute the angles between the tangent lines of the
three walls (see Fig.~\ref{fig:schematic}) and plot the distributions
in the patterned case in Fig.~\ref{fig:mutual}. The parasite-dominated
sectors appear to be rather narrow (typically 60 degrees) in comparison to the
host- and empty-dominated ones, whose typical spread equals approximately
160 degrees. This is somewhat balanced by the larger variance of the width
of the parasite-dominated sectors. In any case, these observations conform 
to those one can make by studying the appearance of Fig.~\ref{fig:domains}
in the vicinity of the vortices by bare eye.

The ratio of the measured domain wall length and its counterpart in random
configurations is plotted in Fig.~\ref{fig:ll} as a function of the two
spreading rate parameters, $\lambda_h$ and $\lambda_p$. The parameter space
is clearly divided into two regions with a continuous crossover in between
by this quantity. In one of these parts, the ratio equals approximately one
(corresponding to homogeneous population densities), whereas in the other, 
the ratio clearly differs from one (corresponding to pattern formation).
In other words, in cases where the system does not exhibit patterns visible 
to human eye, the domain wall length statistics also coincides with that 
in a spatially homogeneous case, whereas with patterns there is a significant 
difference. Note, however, that the numerical value of the ratio bears no 
meaning by itself, since it can be tuned to almost any value by altering the
smoothing kernel (in particular the smoothing width $w$) in the 
coarse-graining procedure. Nevertheless, for any usable value of $w$, the
ratio clearly deviates from one and therefore is a usable metric of the
patterning. The transition between the cases is continuous everywhere on 
the transition line. We have also considered similar statistics for the
average vortex jump length and the average vortex lifetimes, recovering
a continuous crossover in each case.

Given all these results on the vortices, the domain walls, and their dynamics, 
there remains the question of how much of them can be explained by considering 
the vortices as identical random walkers with random signs in a plane. To study
this point in detail, we have performed numerical experiments consisting of a 
set of simulated random walks for a given value of the simulation parameters as
follows. For each walk, draw its duration $N_{rw}$ randomly using the measured 
vortex lifetime distribution. Let each walk start at the origin and perform 
$N_{rw}$ random steps. For each step, draw the jump length from the 
corresponding measured jump length distribution and choose the direction in two
dimensions randomly from a uniform distribution. Record the average total 
length of the walks for each case of interest, i.e.~for different simulation 
parameters (and consequently different measured distributions).

In the non-patterned and patterned cases, we get average total walk lengths
of 20 and 9 lattice units, respectively, in the random walk experiment. In
the non-patterned case, this roughly corresponds to half of the average
domain wall length which, in turn, approximately equals the average 
inter-domain distance. Therefore, the picture of vortices as merely 
uncorrelated random walkers cannot be ignored. On the other hand, there
is more than an order of magnitude of difference between the average domain
wall and random walk lengths. This rules out any explanation of the dynamics
of the system that is based on the random walk picture only.

Finally, we take a look at the behavior of the number of vortices as a 
function of time. A relevant benchmark for comparison is the complex
Ginzburg-Landau equation (CGLE) which has been studied extensively
(for a good review, see \cite{aranson02}). Both in the usual deterministic
version of the equation \cite{gil90}, and in a version with added noise 
\cite{wang04}, the number of pairs of vortices $n(t)$ has been observed
to be a Markov process with $n(t)$-dependent creation and annihilation
rates. An essential feature of such processes is the lack of memory of
any kind. We have partially repeated the analysis of 
Refs.~\cite{gil90,wang04} in the present case. Average creation and
annihilation rates as a function of $n(t)$ can be measured once a 
sufficient number of realizations of the process have been simulated. Doing
this, we arrive at a description similar to Eqs.~(3) and (4) of 
\cite{wang04}. In other words, the creation rate is a constant and
the annihilation $\Xi_{-}(n)$ rate assumes the form
\begin{equation}
\Xi_{-}(n) = An^2 + Bn \, 
\label{eq:markovannihilationrate}
\end{equation} 
where $A$ and $B$ are constants. In this respect the number of vortices
behaves here in the same way as in the CGLE. However, a detailed look
reveals again an important difference. In the present case, the number
of vortices as a function of time contains an oscillatory component
clearly visible both in the time series itself and its Fourier power
spectrum. This cannot be explained by a one-dimensional memoryless
dynamical system, and therefore the vortex number statistics 
lies outside the realm of applicability of the analysis of
Refs.~\cite{gil90,wang04}. 

Instead, to recreate the vortex number process, short-term memory has to be
incorporated in a way or another. One way to do this is to consider a 
two-dimensional process where the role of the second variable is played by
the first discrete time derivative of the vortex number $n$. This leads to
a description similar to Eqs.~(\ref{eq:lin}) and (\ref{eq:matrix_elements}) for
the vortex number and its derivative. On the other hand, the same process can
be formulated in terms of the second time derivative or a temporal delay. In
any case, one-dimensional processes such as 
Eq.~(\ref{eq:markovannihilationrate}) do not suffice. 

\section{Dynamics}

\subsection{Effective mean-field}

\begin{figure}[!h]
\begin{center}
\includegraphics[width = 8cm]{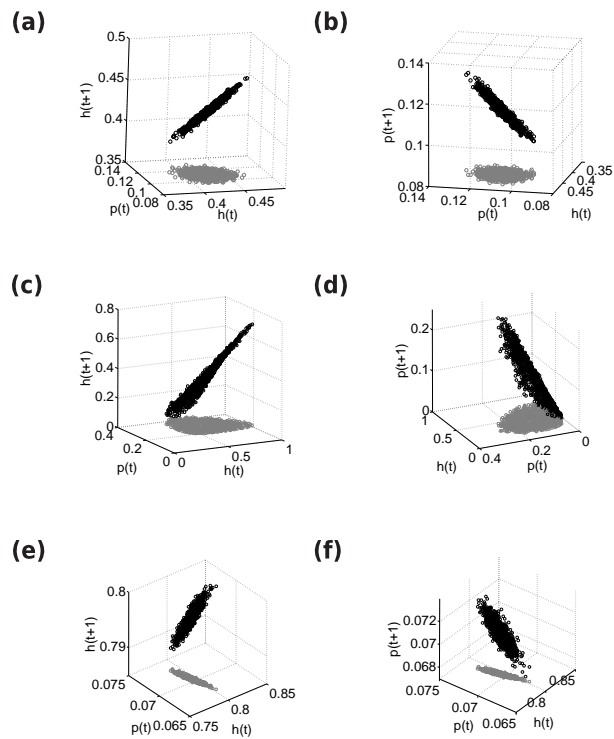}
\end{center}
\caption{The Poincar\'e maps illustrating the dynamics of the system. 
(a), (b): $h_{t+1}$ and $p_{t+1}$ as a function of $h_t$ and $p_t$ for
a system in the patterned state (parameters as in Fig. 1a) with size
$L \times L = 512 \times 512$. (c), (d): The same for a system of size
$64 \times 64$. (e), (f): As in (a) and (b) but for a system in the
homogeneous state, i.e.~with $\lambda_p=1.3$. The simulations have been run
for 3000 time steps and data was gathered for the last 1000.}
\label{fig:poincare}
\end{figure}

To gain insight on the dynamics of the system, and to produce the 
effective low-order corrections to the MF equations, we have used Poincar\'e
maps \cite{zaslavsky} reconstructed from simulations. Here, the dynamical
(scalar) variables of the system at time $t+1$ are considered to be 
functions of the same at time $t$. This mapping is then numerically
computed from measurements of the dynamical variables from the simulations
of the model. In particular, in the case of two dynamical variables, a
graphical representation of the Poincar\'e maps boils down to two 
three-dimensional scatter plots. We draw them for two different system sizes
in the patterned case and for comparison in the non-patterned one in 
Fig.~\ref{fig:poincare}. Immediate conclusions can be made. There is indeed
a clear functional dependence of the densities on their previously assumed
values. In other words, the dynamics can be described by using the 
densities themselves only; as a dynamical system the present one is
two-dimensional. This would not have been obvious given the $2L^2$ discrete
degrees of freedom in the spatially extended system. 

We have also verified this conclusion using attractor reconstruction 
\cite{packard80}. There, a simulated time series is studied using a variable
number $N_d$ of degrees of freedom, which are the time series itself as such,
and with suitable delays $\tau$, $2\tau$, $3\tau$, and so on. Here, we have 
used a quarter of the time of a period as the time delay $\tau$. With two 
dynamical variables, $h(t)$ and $h(t-\tau)$ we find that there is a unique
mapping (up to noise) from their values at time $t$ to their values at time
$t+1$. Such mapping does not exist if only one dynamical variable is used, 
and these findings hold also when $p(t)$ (and the corresponding delayed 
variable) is used instead of $h(t)$. Therefore, attractor reconstruction
duplicates the conclusion from above that the system has two degrees of 
freedom as a macroscopic dynamical system. 

Based on the numerical observations, the dynamics can be cast as an
analogous pair of equations to (\ref{eq:naivemf}) but with different functional
forms. Even further, the plots are clearly linear in both variables for a 
large system size (here $L=512$, Figs.~\ref{fig:poincare}a and 
\ref{fig:poincare}b), whereas for smaller system sizes ($L=64$, 
Figs.~\ref{fig:poincare}c and \ref{fig:poincare}d) there are deviations from
the linear form. The point clouds are stretched towards the absorbing state
of the system with extinct parasitoids. The necessary system size for the
onset of linearity can be crudely estimated by visual inspection. It reveals 
the rough approximation $L_{\mathrm{lin}} \approx 120$, slightly larger than
the corresponding average domain wall length (see Sec.~\ref{sec:lines}). Thus,
the linearity is obeyed, if the system is large enough to accommodate several
domains.

To arrive at an effective MF iteration based on the Poincar\'e plots
valid for large systems, we perform least-squares fits to them and
arrive at Eq.~(\ref{eq:lin}) with numerically determined coefficients
$a_{\sigma,\sigma'}$. This picture takes space implicitly into account
since the interaction parameters (or the matrix elements) are
automatically suitably adapted to the form and parametrization of the
manifolds. In addition to this deterministic description, the system 
exhibits noise whose magnitude -- both perpendicular and tangential to 
the manifolds -- is proportional to the square root of the system size. Also,
after the pure noise component and the deterministic picture above have been
removed, one is left with a residue that comes from the approximation made in
the linearization. However, for large system sizes this residue is small when
compared to the linear deterministic component and the noise. 
The numerically fitted coefficients $a_{\sigma, \sigma'}$ come into play
below during the discussion of the effective parameters and sustained
oscillations.

\begin{figure}[!h]
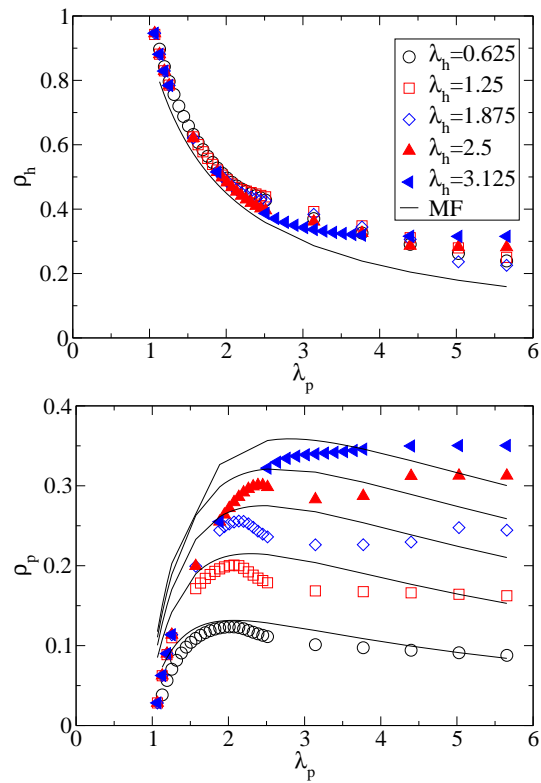

\begin{center}
\includegraphics[width = 7cm]{fig10a.eps}\\
\includegraphics[width = 7cm]{fig10b.eps}\\
\end{center}
\caption{Prevalences as functions of the spreading rate 
parameters $\lambda_{h|p}$, together with the MF predictions. The
prediction for $\rho_h$ does not depend on $\lambda_h$, whereas the 
for $\rho_p$ predictions for 
$\lambda_h=0.625, 1.25, 1.875, 2.5$ and $3.125$ 
from bottom to top are shown. The color code is the same in both panels.}
\label{fig:prev}
\end{figure}

The predictions of the MF theory of Eqs.~(\ref{eq:naivemf}) for the host and 
parasitoids densities compared with those obtained from simulations are shown 
in Fig.~\ref{fig:prev}. The host density $\rho_h$ does not depend on the host 
spreading rate parameter $\lambda_h$ according to the MF theory and as is seen 
from the figure, this is almost the case also in simulations. On the other 
hand, the parasitoid density $\rho_p$ depends strongly on both spreading rate 
parameters $\lambda_h$ and $\lambda_p$ both in the MF theory and in the 
simulations. Altogether, Fig.~\ref{fig:prev} shows that the MF approximation 
works surprisingly well in predicting the species densities. Note however, that
these comparisons are only for the average densities and do not constitute 
evidence of the validity of the MF theory in other respects, such as the 
stability of the fixed point or the nature of oscillations, if any. 

\begin{figure}[!h]
\begin{center}
\includegraphics[width = 4cm]{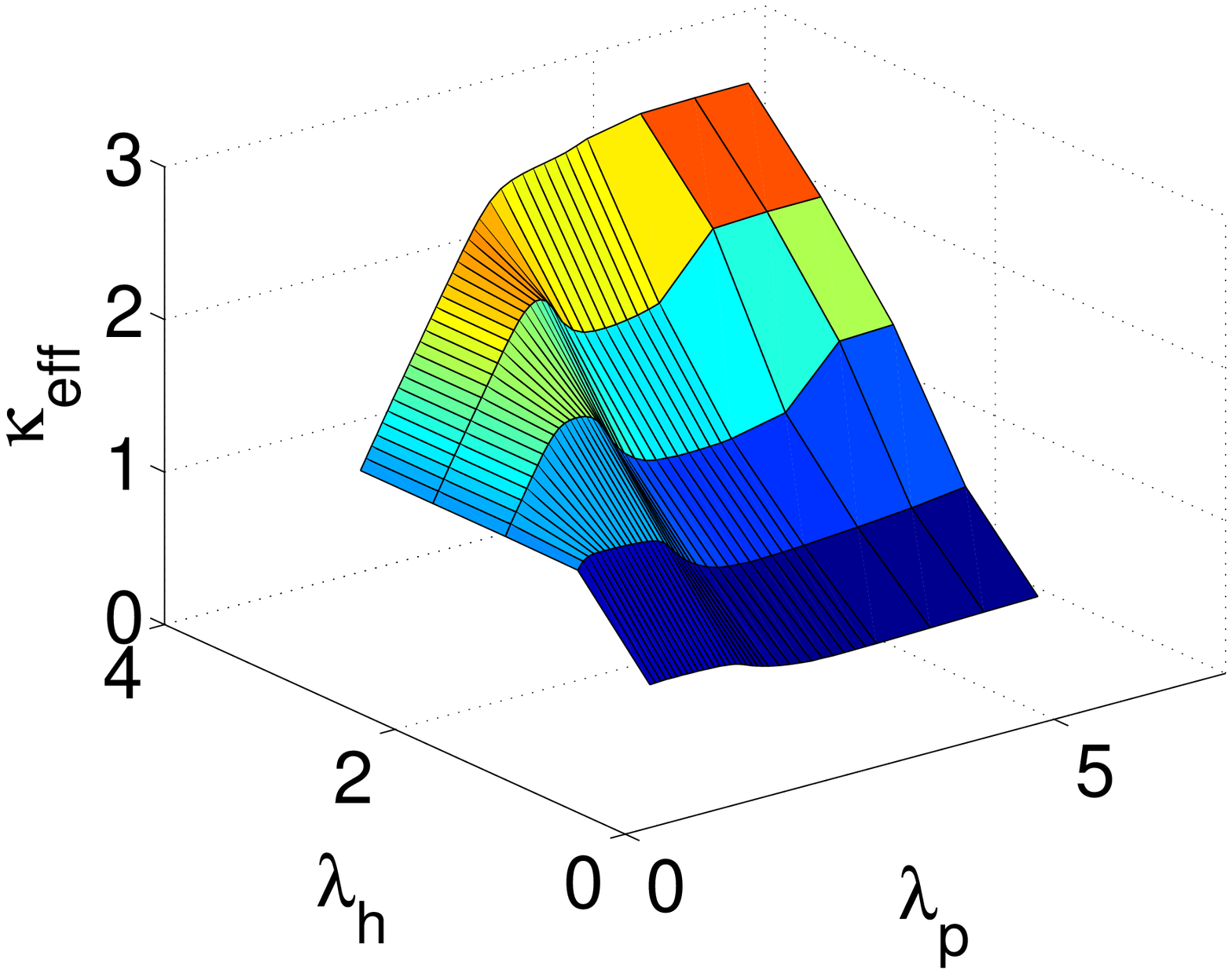}
\includegraphics[width = 4cm]{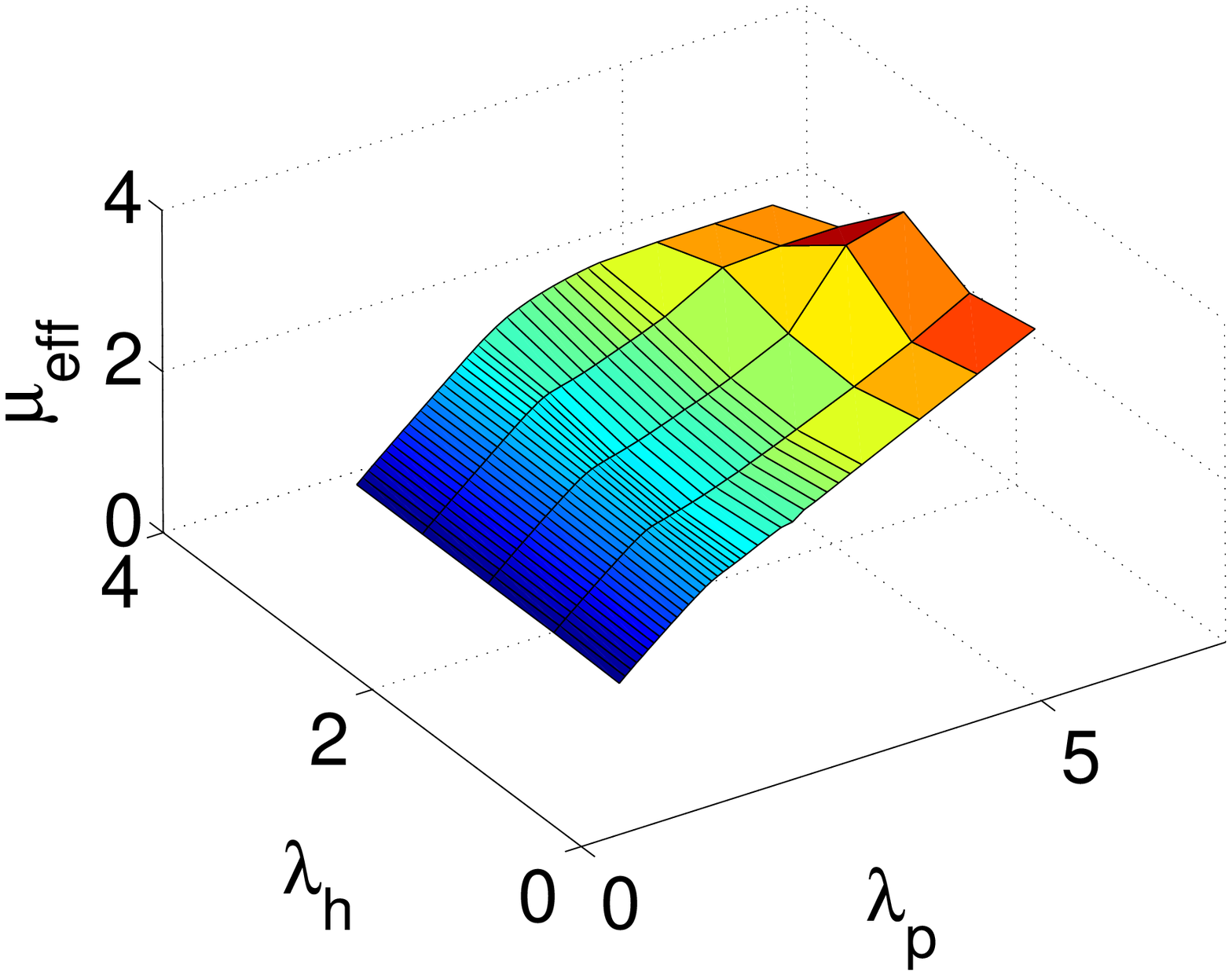}
\end{center}
\caption{Effective values of the parameters $\kappa$ (left)
and $\mu$ (right) as a function of the input parameters
$\lambda_h$ and $\lambda_p$.}
\label{fig:effparam}
\end{figure}

In order to arrive at equations that can also predict these more complicated
issues, we proceed in two phases. The first one -- here called the zeroth-order
correction -- is to ask which values of the parameters have to be inserted into
the MF equations so that they give the same population densities as the 
simulations. In other words, the question is what are the effective MF 
spreading rate parameters as a function of the simulational ones. To this end, 
we have run simulations for a wide range of parameters and computed the
effective host and parasitoid spreading rates $\kappa_{\mathrm{eff}}$ and
$\mu_{\mathrm{eff}}$ by using the inverse of Eq.~(\ref{eq:nontrivialfp}). The 
results are plotted in Fig.~\ref{fig:effparam}. The conclusions are evident. 
The dependence is strongly nonlinear, as was expected, and the immediate 
observation is therefore that the effective parameters cannot be obtained from
the simulational ones by any simple rescaling. Furthermore, the division 
between the patterned and non-patterned regions of the parameter space is 
clearly visible in Fig.~\ref{fig:effparam}. It shows up as a ridge in 
$\kappa_{\mathrm{eff}}$ and as a bend in $\mu_{\mathrm{eff}}$. 

In addition to what has been discussed above, the effective parameters do 
depend on the population densities within single realizations with fixed 
parameters. Here, we call this dependence the first-order correction, and it is
the one that causes changes in the effective iteration equations with respect 
to Eqs.~(\ref{eq:naivemf}). Indeed, given the described dependence, $\kappa$ 
and $\mu$ in Eqs.~(\ref{eq:naivemf}) are functions of $h_t$ and $p_t$ as 
opposed to constants. To compute the first-order correction, linearize 
Eqs.~(\ref{eq:naivemf}) around the reactive fixed point assuming that $\kappa$ 
and $\mu$ are functions of the population densities. One arrives at 
Eq.~(\ref{eq:lin}), in which there are four matrix elements that are linear in 
the four unknown first derivatives of $\kappa$ and $\mu$ with respect to $h_t$ 
and $p_t$. The matrix elements are then set to equal those obtained from linear
fits to the Poincar\'e plots, i.e.~the $a_{\sigma,\sigma'}$ in 
Eq.~(\ref{eq:lin}), and the unknown derivatives are solved for. 

\begin{figure}[!h]
\begin{center}
\includegraphics[width = 7cm]{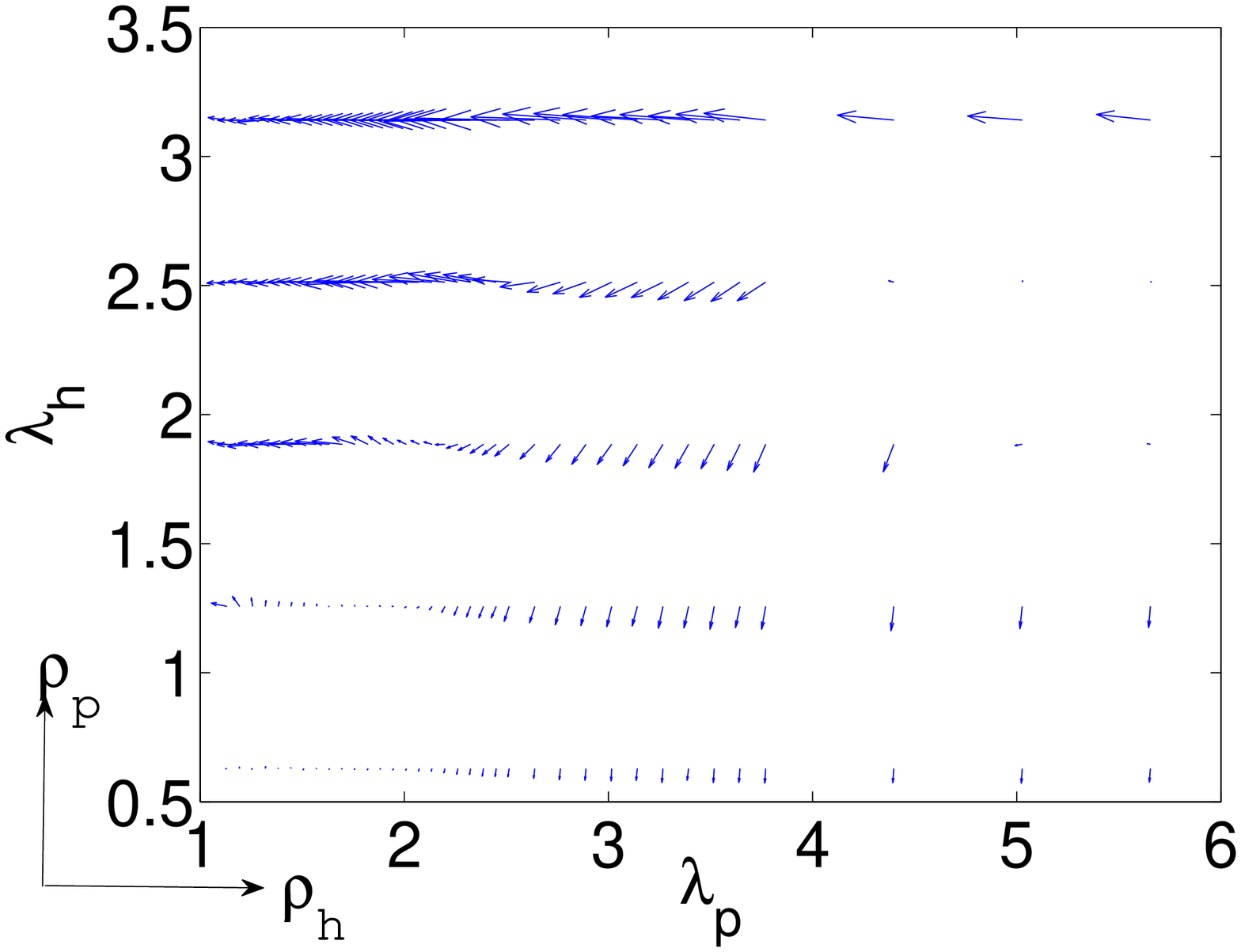}\\
\includegraphics[width = 7cm]{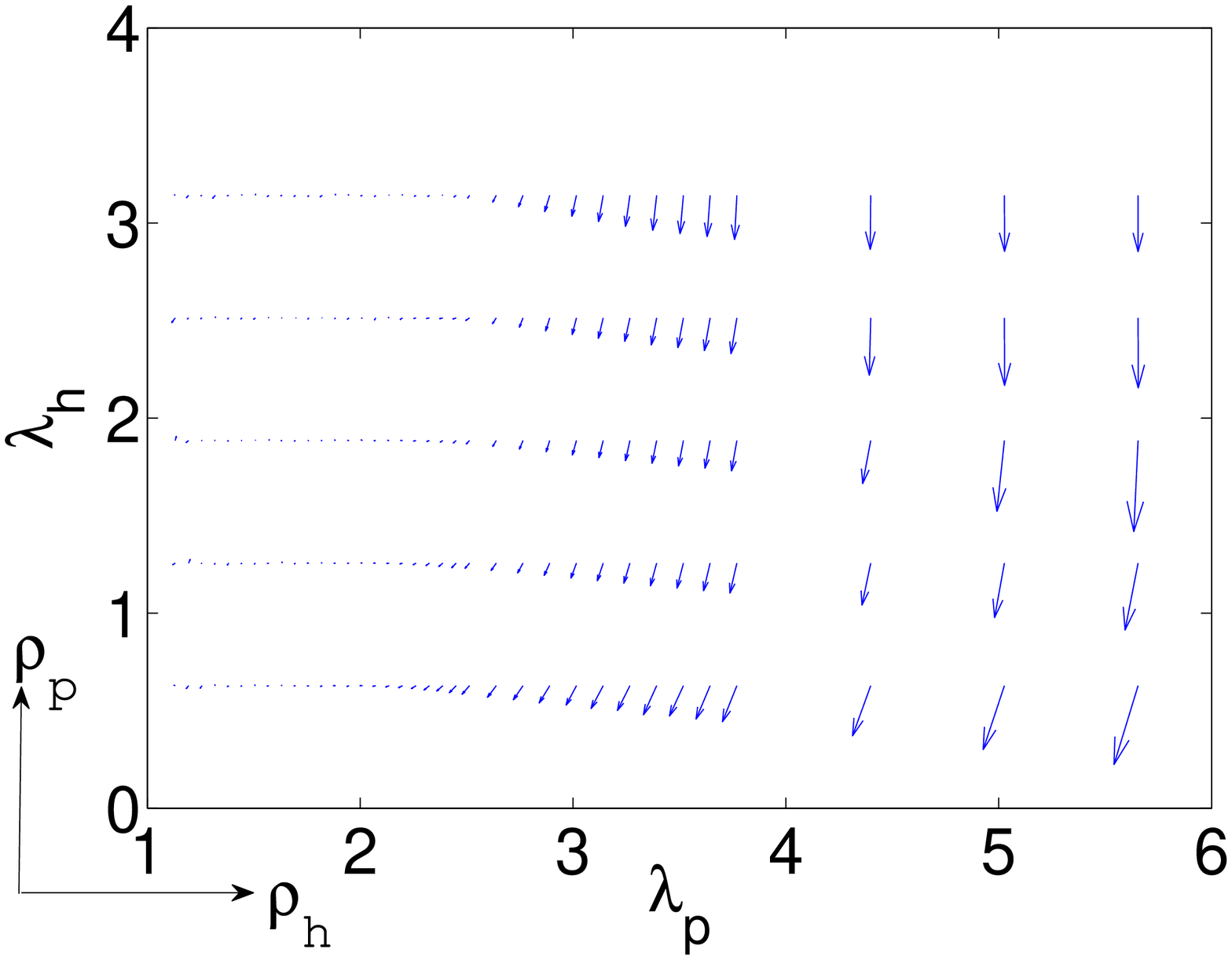}
\end{center}
\caption{The first-order dependence of the
effective parameters on the population densities $\rho_h$ and $\rho_p$. 
At each point, the arrow depicts the derivative of $\kappa_{\mathrm{eff}}$ 
(above) and $\mu_{\mathrm{eff}}$ (below) according to the auxiliary coordinate 
system drawn on bottom left. The same points of the parameter space are sampled
as in Fig.~\ref{fig:effparam}.}
\label{fig:effparamquiver}
\end{figure}

The results are shown in Fig.~\ref{fig:effparamquiver}, in which for each 
considered value of the parameters $\lambda_h$ and $\lambda_p$ the derivatives 
are drawn as an arrow depicting the magnitude and direction of the gradient of 
$\kappa$ and $\mu$ with respect to $h_t$ and $p_t$. The auxiliary coordinate 
system on the bottom left corner fixes the orientation of the gradients. The 
behavior of $\mu_{\mathrm{eff}}$ is especially interesting. In the 
non-patterned case the derivatives practically vanish. This means that the rate
parameters $\kappa$ and $\mu$ are effectively constants and that the MF theory 
is a rather good approximation. This, in turn, means that the populations are 
well-mixed. On the other hand, the derivative of $\mu_{\mathrm{eff}}$ with 
respect to $p_t$ clearly deviates from zero and assumes negative values in the 
patterned case. This is a direct sign of the insufficiency of the MF theory in 
this parameter regime. The correlations that for a larger number of 
parasitoids, makes them aggregated within themselves and thus decreases the 
``free boundary'' available for spreading constitute one contributing 
factor to this insufficiency. 

We have also checked the consistency of the computations above by solving
for the effective $\kappa$ and $\mu$ from the MF iteration equations
(\ref{eq:naivemf}) separately for each time step, and considered these
as a function of the population densities. The results are in an excellent
numerical agreement with those above.
In principle, the fits of the functional form in Eq.~(\ref{eq:lin})
could be extended to any order, and used together with the higher-order
counterparts of the analysis of the derivatives of $\kappa$ and $\mu$ above
to extract the derivatives up to any order. 

\subsection{Oscillations}
\label{sec:oscillations}

\begin{figure}[!h]
\begin{center}
\includegraphics[width = 7cm]{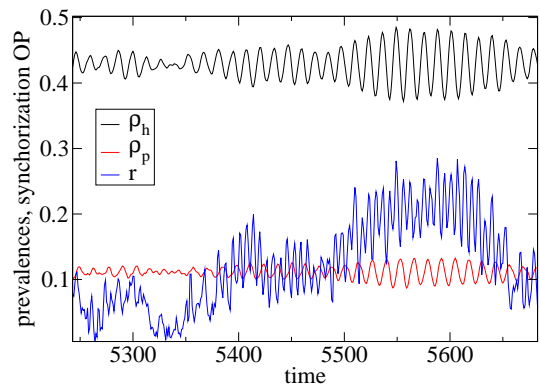}
\end{center}
\caption{Time series of host and parasite prevalences, and the 
synchronization order parameter $r$ (Eq.~(\ref{eq:synchro_op})). The parameters
are as in Fig.~\ref{fig:example}. The amplitudes of the oscillations correlate 
strongly with each other and with the values of $r$. See 
Fig.~\ref{fig:example_timeser} for the dependence of the amplitudes on the 
system size.}
\label{fig:timeser}
\end{figure}

\begin{figure}[!h]
\begin{center}
\includegraphics[width = 7cm]{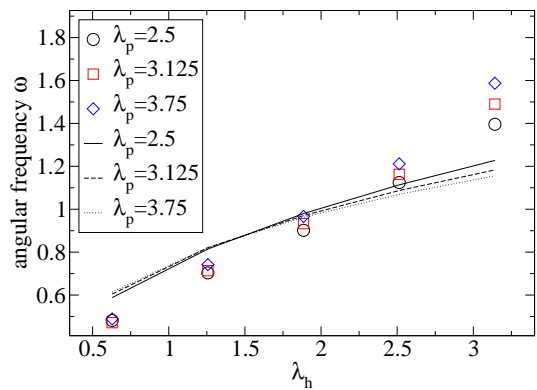}
\end{center}
\caption{The angular frequency of the sustained oscillations as 
a function
of the host spreading rate parameter $\lambda_h$ for several values of 
$\lambda_p$. The frequency is measured as the polar angle of the eigenvalues
of the linear fit of the simulation data to Eq.~(\ref{eq:lin}). 
The measurements
are done for a system of size $L \times L = 512 \times 512$ which is large 
enough for the linear fits to work well. Other parameters are as in 
Fig.~\ref{fig:example}. The solid, dashed, and dotted lines show the frequency
predicted by MF theory, i.e.~linearization of Eq.~(\ref{eq:naivemf})
and numerical solution of the resulting eigensystem.}
\label{fig:frequency}
\end{figure}

With patterns the population densities oscillate with an amplitude that
fluctuates in time (Fig.~\ref{fig:timeser}). Without patterns, the 
densities fluctuate randomly around a stable fixed point. The angular
frequency of the oscillation can be measured by performing 
three-dimensional linear fits to the Poincar\'e maps (Fig.~\ref{fig:poincare})
and extracting the imaginary part of the eigenvalues of the resulting
matrix $a_{\sigma,\sigma'}$ in Eq.~(\ref{eq:lin}).
The MF prediction for the frequency is computed
similarly. These are compared in Fig.~\ref{fig:frequency}. The frequencies 
from the MF theory and the simulations have roughly the same numerical 
values, but the agreement is far from perfect. 

Instead of the oscillations themselves, 
the most characteristic property of the
population densities is the fluctuation of the amplitude. The typical 
amplitude depends both on the parameters and the system size (see 
Fig.~\ref{fig:example_timeser}). Decreasing the system size increases
it but keeps the qualitative behavior and
the population densities intact. We have previously given an explanation
for the amplitude fluctuations \cite{peltomaki08} which we here briefly
rephrase. Namely, both with and without patterns, the fixed point of
the effective (deterministic) iteration given by the fits to the Poincar\'e
maps is stable. As such, there should be no oscillations in either case. 
However, the eigenvalues of the iteration matrix are a complex-conjugated
pair in both cases, and the transients towards the fixed point are
oscillatory. Such transients give naturally rise to two independent time
scales, one for the oscillation (time of a period) and another one for
the decay. Denoting the eigenvalue pair as $\rho e^{i\phi}$, the ratio of
the time scales is
\begin{equation} \label{eq:timescale_ratio}
\nu = \frac{\tau_{\mathrm{decay}}}{\tau_{\mathrm{osc}}} 
= \frac{\phi}{\log \rho} \, .
\end{equation}
Inspecting this numerically reveals small values without patterns and
large values with them. Therefore, in the latter case any fluctuations
away from the fixed point lead to slow oscillatory convergence towards 
the fixed point which is repeatedly disrupted by stochasticity giving
rise to a perpetual decay -- noise-sustained oscillations.

Note that when approaching the stability limit from below, 
$\rho$ approaches one, and the ratio $\nu$ diverges towards positive infinity.
With $\rho > 1$, the ratio reassumes finite values, now with a negative sign,
but does not anymore carry the same physical meaning as below the limit. 

\begin{figure}[!h]
\begin{center}
\includegraphics[width = 7cm]{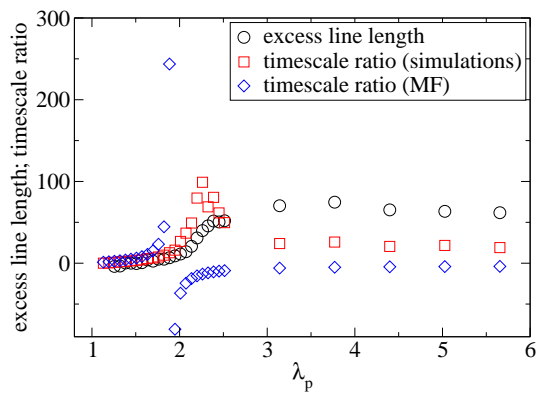}
\end{center}
\caption{The timescale ratio 
of Eq.~(\ref{eq:timescale_ratio})
in both the simulations and the MF theory plotted together with
the difference between the average domain wall length in the actual
simulation and in random configurations. All other parameters except for
$\lambda_p$ are as in Fig.~\ref{fig:example}.}
\label{fig:timescales}
\end{figure}

The time scale ratio $\nu$ both measured from simulations and predicted by the 
MF equations and the excess domain wall length (the difference between the 
average domain wall length and its value in random configurations) are plotted 
in Fig.~\ref{fig:timescales} as a function of the parasitoid spreading rate 
parameter $\lambda_p$. The behavior of $\nu$ is drastically different between
simulations and theory. In the latter case, the ratio diverges at around 
$\lambda_p=1.9$ and assumes negative values at higher $\lambda_p$. In other 
words, the system undergoes a bifurcation from a stable fixed point to an 
unstable one characterized by a limit cycle. On the other hand, the simulated 
time scale ratio $\nu$ peaks around almost the same point, but does not in fact
diverge, signaling that there is no limit cycle involved. To see the connection
between the temporal and spatial properties of the system, note that the 
dependence of the average domain wall length and the time scale ratio on the 
parasitoid spreading rate parameter $\lambda_p$ is similar for 
$\lambda_p < 2.2$. This can be interpreted as direct evidence that the pattern 
formation and the deviations of the dynamics from the MF description are 
closely intertwined. Also the behavior of the time scale ratio and the excess 
line length is qualitatively similar for $\lambda_p > 2.2$ since both are 
constants in this regime. This is comparable to the coupled discrete 
oscillators of Wood and co-workers \cite{wood06prl, wood06}, where a related 
system has a second-order synchronization transition in the MF theory and in 
higher dimensions but not in two.

There is also another straightforward spatial interpretation of the amplitude
fluctuations. Consider the system divided into small boxes whose width is 
clearly smaller than the typical width of the coarse-grained  patterns, but 
which at the same time are large enough so that local self-averaging occurs 
inside each box. Now a local phase $\phi_i$ for each box $i$ can be defined as 
the phase angle of the population densities inside the box with respect to the 
average densities (see Fig.~\ref{fig:domains}). Given these, the degree of 
synchronization between the boxes can be defined as is customary in the context
of the Kuramoto model for coupled oscillators
\cite{acebron05}
\begin{equation} \label{eq:synchro_op}
r = \frac{1}{N}|\sum_j e^{i \phi_j}| \, ,
\end{equation}
where the summation runs over all $N$ boxes.

In this computation, we have used the box size $l \times l = 64 \times 64$ in 
accordance with the criteria above, and computed the synchronization order 
parameter $r$ for each time step. This is plotted in Fig.~\ref{fig:timeser}
together with the host and parasitoid densities in the same simulation run. 
The amplitude of the density oscillations correlates strongly with $r$. This is
to be interpreted such that the system spontaneously synchronizes over at least
intermediate length scales at random intervals. As a feature of secondary 
importance, the synchronization order parameter $r$ oscillates with a time of a
period half of that of the population densities, which is also seen in the 
Fourier power spectrum of $r$ (not shown).

\subsection{Phase diagram}
\label{sec:results_phasediagram}

\begin{figure}[!h]
\begin{center}
\includegraphics[width = 7cm]{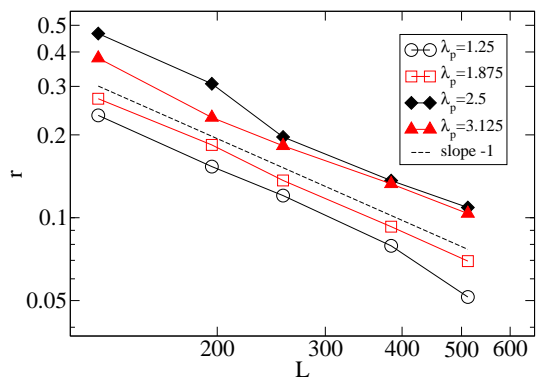}
\end{center}
\caption{The synchronization order parameter $r$ of 
Eq.~(\ref{eq:synchro_op}) as a function of system size $L$ for several
values of $\lambda_p$. In each case, the scaling $r \sim L^{-1}$ is
found, demonstrating the lack of long-range order in the system.}
\label{fig:synchro_op_vs_L}
\end{figure}

Given the fact that the noise-sustained oscillations can be interpreted in 
terms of spontaneous synchronization, it is an interesting question whether it 
is macroscopic in character or ``merely'' a finite-size effect. To address this
 issue, we have computed the temporal average of the synchronization order 
parameter $r$ (\ref{eq:synchro_op}) for different system sizes $L$ both
in the patterned and non-patterned cases. The results are shown in
Fig.~\ref{fig:synchro_op_vs_L}. In all cases, the scaling $r \sim L^{-1}$ is 
found (similarly to \cite{wood06prl,wood06}). Thus, we conclude that the system
does not show long-range order and thus does not completely synchronize at 
macroscopic scales. 

In spite of this, the phase diagram remains interesting. In addition to the 
extinction transition of the parasitoids, the coexistence phase can be split 
into three. There are the non-patterned and the patterned regions, and since 
the crossover between them is not sharp a ``gray area'' between them. We have 
built the phase diagram in the $(\lambda_h,\lambda_p)$-plane by defining the 
boundary between the non-patterned case and the gray area to be the line at 
which the timescale ratio $\nu$ of Eq.~(\ref{eq:timescale_ratio}) equals 1.0, 
and that of the gray area and the patterned phase where the ratio equals 4.0.
Albeit arbitrary, these definitions seem fit since upon looking at the 
population densities as a function of time and figures such as 
Fig.~\ref{fig:example}, the patterns and the oscillations are clearly not there
if the ratio is below one, and, on the other hand, if it exceeds 4, both are 
clearly observed. Note that the timescale ratio assumes values as high as 100 
(see Fig.~\ref{fig:timescales}). The qualitative features of the phase diagrams
are not sensitive to altering the limiting values of $\nu$.

\begin{figure}[!h]
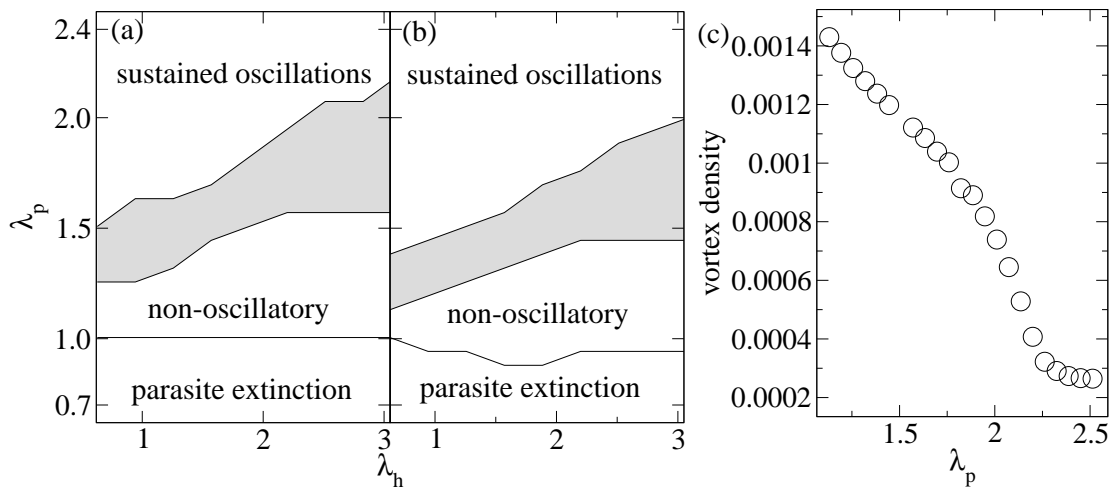

\begin{center}
\includegraphics[width = 9cm]{fig17.eps}
\includegraphics[width = 5.5cm]{fig17c.eps}
\end{center}
\caption{Phase diagrams with varying $\lambda_h$ and $\lambda_p$, for
  different spreading ranges. (a) $w_h \! = \! 3$ and $w_p \! = \!
  1.5$, (b) switched $w_h \! = \! 1.5$ and $w_p \! = \! 3$. 
  In both cases four ``phases'' are found, parasite extinction, coexistence
  without oscillations, coexistence with oscillations, and a gray area between
  the two, which is due to the fact the there is not, in fact, a sharp
  transition between the oscillatory and non-oscillatory regimes, but a
  continuous cross-over. The upper and lower bounds of the gray 
  area are defined
  as the lines where the timescale ratio of Eq.~(\ref{eq:timescale_ratio})
  equals 1 and 4, respectively. The system size is 
  $L \times L = 512 \times 512$. (c) The vortex density as a function of 
  the parasitoid spreading rate parameter $\lambda_p$ for the case of panel (a)
  with $\lambda_h=0.63$ and the smoothing width $\sigma=8$.}
\label{fig:phases}
\end{figure}

The phase diagrams are shown in Fig.~\ref{fig:phases} for two choices of the 
spreading widths $w_h$ and $w_p$. It is seen that for any value of $\lambda_h$ 
the four phases (parasitoid extinction, homogeneous coexistence, the gray 
area, and patterned coexistence) follow each other in this order if the 
parasite spreading rate parameter $\lambda_p$ is increased. This feature does 
not depend on the particular values of the spreading widths, or even their 
mutual order. The phase structure bears some resemblance to related earlier 
models \cite{satulovsky94,tome07,arashiro08} in that oscillatory and 
non-oscillatory regions and a boundary between them is recovered. See 
Sec.~\ref{sec:discussion}  for a detailed discussion and comparison. 
Fig.~\ref{fig:phases}c also shows the vortex density, i.e.~number of vortices
per lattice site, as a function of $\lambda_p$. One sees that the number of 
vortices decreases rapidly while going from the nonoscillatory region via the
gray area to the oscillatory regime. 

Finally, we note that even if the system does not in general synchronize
macroscopically, such synchronization can be artificially achieved by making
the system small with respect to the spreading widths. This can originate in 
either making the system really small, or keeping the size fixed and 
increasing the spreading widths. In these cases, the MF theory probably works
better than in what has been studied here. This, however, lies outside the
scope of interest since extremely small system sizes or very large spreading
lengths are needed.

\section{A realistic example}

To give a concrete example of how to apply the methods introduced here, 
we illustrate the role of coherent dynamics in biological systems by
resorting to a well-known metapopulation system, investigated at
length by the Metapopulation Research Group at University
of Helsinki, Finland 
\cite{hanski98,nouhuys02,nouhuys04,hanski94,ovaskainen01}. The species
in question are the Glanville fritillary butterfly \emph{Melitaea
cinxia} and its specialist parasitoid wasp \emph{Cotesia
melitaearum} on the network of habitats on the \AA land islands in
the Baltic Sea (60$^\circ$ N, 20$^\circ$ E) between Finland and Sweden. 
Here we use the known patch
geography as the ``lattice''. The two populations
follow a variant of the model used above, with the differences that
the contribution of each patch to the connectivities is multiplied by its
area, and that in addition to the connectivity-driven spreading a small
amount of uniformly random reproduction is used. These differences are
biologically motivated. Intuitively, the immigration pressure
from a given patch increases with the local population, which itself
is positively coupled to the amount of locally available suitable habitat,
the patch size. The random spreading mimics occasional long-range
dispersal, and its effect is to prevent extinction on remote subnetworks,
which are naturally present at the outskirts of the sparse archipelago. 
Also, open boundaries are used. This does not restrict the validity of the
analysis for two reasons. First, the equations (\ref{eq:lin})
are an observation
made {\it from} the simulations, and by plotting the Poincar\'e maps as in 
Fig.~\ref{fig:poincare}, one sees that the equations hold both for open and 
periodic boundaries. Second, the equations (\ref{eq:matrix_elements}) are 
obtained from Eqs.~(\ref{eq:generalmf}) by straightforward differentiation, 
and Eqs.~(\ref{eq:LV}), in turn, are a rough approximation equally valid in 
either case.

\begin{figure}[!h]
\begin{center}
\includegraphics[width = 10cm]{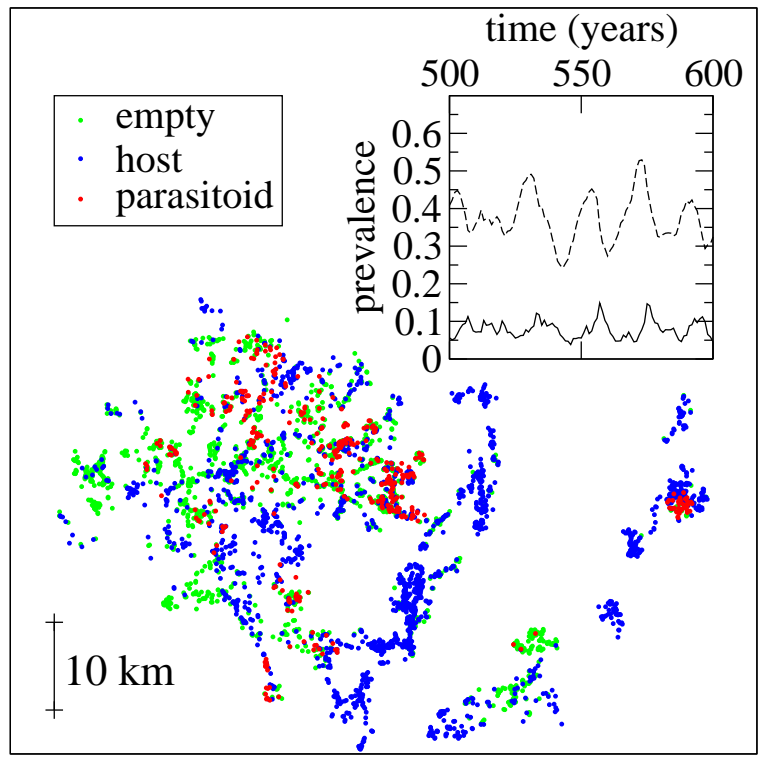}
\end{center}
\caption{Main figure: a snapshot of the simulated dynamics of hosts
  and parasitoids on the \AA land patch geometry. Each dot corresponds
  to a single patch. Inset: the prevalence of   the hosts (dashed
  line) and parasitoids (solid line) as a function of time. As
  spreading widths we have used $w_h=1000$ m, $w_p=500$ m. The annual
  death probability is $\delta=0.9$ and the host and parasitoid
  spreading rate parameters are $\lambda_h=100$ and
  $\lambda_p=1400$. A fraction of 0.01 of the hosts spreads randomly
  to all patches. The archipelago is roughly 60 km $\times$ 80 km in
  size.} 
\label{fig:alandexample}
\end{figure}

Fig.~\ref{fig:alandexample} shows a snapshot of the system,
and the inset contains the respective time series; see the caption for
the parameters. The question is now, do we find evidence of
oscillations and/or patterns here? A similar analysis of the dynamics, from
Eq.~(\ref{eq:lin}) is readily carried out, revealing the
eigenvalues $\lambda_\pm \approx 0.84 \pm 0.17i$ for the parameters
used in Fig.~\ref{fig:alandexample}. These imply, in accordance with
the appearance of the time series, that the system contains patterns
and exhibits oscillations coming from the combination of an 
unstable fixed point with oscillatory transients
and demographic stochasticity. The patterning is clearly visible in
Fig.~\ref{fig:alandexample}. It can be analyzed using the smoothed densities
by considering the sum in the expression for them, 
Eq.~(\ref{eq:smoothed_densities}), as a summation over the patches, and using 
the position-dependent smoothed densities assuming average population densities
at each patch as the triple point in the phase space 
(see Fig.~\ref{fig:domains}a). The rest of the analysis can then be carried out
as explained in Section \ref{sec:characterizing_patterns}. Doing this, one 
finds a domain structure analogous to that on the regular lattice with the 
average domain wall length 
$\langle \ell \rangle/\langle \ell_{\rm random} \rangle \approx 1.8$.

These observations are of interest for a couple of reasons. First, the \AA land
landscape is rather heterogeneous, and at a closer inspection \cite{vuorinen04}
appears to consist of a variety of densely connected subsystems, which are 
weakly coupled to each other. Due to the differences in the landscape with 
respect to a regular lattice, it is not clear, a priori, that the whole system 
has similar pattern-included noise-sustained oscillations as the version 
defined on a regular lattice. Instead, the expectation for a single densely 
connected subsystem is to be closer to the mean-field limit \cite{peltomaki05} 
obeyed by a fully connected (essentially non-spatial) system, and since the 
number of such dense subsystems is relatively small, the expectation is the 
same for the full system. However, noise-sustained oscillations are observed, 
and the \AA land system works as an example of the applicability of the 
analysis presented here to more realistic cases, than regular lattices. This is
emphasized by the fact that also the dynamics of the model contains differences
with respect to the lattice one, in addition to the spatial structure. In any 
case, the eigenvalues of the effective iterative map (Eq.~(\ref{eq:lin})) 
provides information about the stability properties and the level of patterning
in the system. 

\section{Discussion}
\label{sec:discussion}

The change between the patterned and non-patterned cases is rather abrupt in
terms of several quantities. This leads naturally to the question whether the
phenomenon is actually a phase transition. We argue that this is not the case
due to several reasons. First, since phase transitions are defined only for 
infinite systems, the transition in plots such as Fig.~\ref{fig:ll} should 
become sharper as the system size increases. However the behavior remains
essentially the same as in Fig.~\ref{fig:ll} for a wide range of system sizes,
except for the fact that for the smallest systems, the highest value the 
average domain wall length assumes becomes smaller due to finite-size effects. 
In any case, no sharpening of the transition is observed, and in this respect 
the correct description of the behavior is as a continuous crossover. 

Second, similar models have been studied with the transition aspect in mind. 
The most relevant one for the present discussion is that by Wood and coworkers
\cite{wood06prl,wood06}, which deals with a symmetric three-state model of 
discrete coupled oscillators. It has been found that the system does not have
a phase transition in two dimensions but does in three and higher dimensions. 
However, in two dimensions, there is a continuous change in the order parameter
around the region of the parameter space where the transition would take place 
in higher dimensions. This is similar to what has been observed here for the 
average domain wall lengths, giving further credibility for the interpretation
of the change from non-oscillatory to oscillatory behavior as a continuous
cross-over. 

The eigenvalue structure in the non-patterned case also speaks in favor of
the absence of a sharp transition. We get a complex conjugate eigenvalue pair
for the matrix $a_{\sigma,\sigma'}$ in Eq.~(\ref{eq:lin}) even in 
the non-patterned case. The difference lies in the relative magnitudes of the 
oscillation and decay time scales. So, even if the ratio of the timescales 
were used instead of the average domain wall length, there would be no sharp 
transition (cf.~Fig.~\ref{fig:timescales}). This can be viewed as a 
consequence of the corrected MF approach yielding a stable fixed point both
in the patterned and the non-patterned regimes. As a conclusion, the 
difference between the two is strictly speaking only quantitative, or -- put
in other words -- in the non-patterned case the spatial correlations are much
more suppressed than in the patterned one. However, on the other hand, the
quantitative difference between the cases is rather huge: There are practically
no patterns in the non-oscillatory systems as can be seen by looking at the
system itself (Fig.~\ref{fig:example}), the average domain wall length
(Fig.~\ref{fig:ll}), and the random walk experiments outlined in 
Section \ref{sec:effmf}, for example. Therefore, discussing these two regimes
as different appears justified. 

In addition to the discrete-space discrete-time setting studied here, the 
methods of analysis and characterization of patterns and dynamics are equally 
applicable to systems with continuous time or space, for example that in 
Ref.~\cite{ovaskainen06}. Applying the methodology to such systems might reveal
differences between them and their discrete-space counterparts. Studies of 
these would be interesting already as such. One possible cause of the 
differences is that the discretization of space into lattice cells often comes 
with implicit locally density-dependent establishment, i.e.~the local 
population density is restricted such that only one individual can be present 
at one site at a time. Continuous-space approaches typically do not have such 
restrictions unless explicitly included.

A very similar system restricted to nearest-neighbour spreading has been
recently studied \cite{satulovsky94,tome07,arashiro08} both analytically using
the pair approximation and numerically. The authors of these articles have 
found that the co-existence phase is divided into two regions. These are 
oscillatory and non-oscillatory coexistence. The similarity to the present case
is striking. It has been concluded in \cite{satulovsky94,tome07,arashiro08} 
from the pair approximation that the oscillations are a limit cycle, and 
indirectly that, since there are oscillations also in the spatially extended 
counterpart, they have to be limit cycles as well. Comparing to the present 
results reveals that such conclusions cannot be made without further evidence; 
the possibility of noise-sustained oscillations remains, and for small systems 
even that of two kinds of oscillatory regimes. While further studies would be 
needed to resolve this issue, note that in some cases the amplitude 
fluctuates in a way that appears similar to the fluctuations in the present
case. One example is given by Fig.~9a of Ref.~\cite{arashiro08}. On the other 
hand, there are established results on two-dimensional three-state dynamic 
lattice models with limit cycle oscillations \cite{tsekouras06}. Furthermore, 
processes on excitable media have been recently characterized in terms of 
prey--predator systems \cite{otani08}, and the machinery introduced here could 
provide tools for studying those as well.

Other comparable approaches taken recently include Refs.~\cite{abta07,abta07b}.
In these works, a simplified model of two connected patches is studied, and
the oscillations are made dependent either on the phase or the amplitude in
the phase space. Both lead to stabilized oscillations. However, 
Eq.~(\ref{eq:lin}) explicitly forbids such dependencies in the present case, 
and therefore the mechanisms suggested in \cite{abta07,abta07b} cannot be the 
root cause of the stability in the spatially extended case studied here.

Another interesting approach is to map related population models
\cite{reichenbach07,reichenbach08} to the complex Ginzburg-Landau equation
\cite{aranson02}. The mappings rely on the existence of an unstable fixed point
\cite{peltomaki08rps}, and the oscillations are classical limit cycles. 
Therefore, these studies and the present one can be described as 
complementary approaches to each other. 

While the present model is without doubt castable as a partial differential
equation at least in an approximative manner, it is an open question of what
the outcome from such equations would be. Since the fully-mixed approximation
can have both a stable and an unstable fixed point, a mapping to the CGLE
along the lines of \cite{reichenbach08} would not work, and further 
complications are also caused by the highly asymmetric reaction rates 
\cite{peltomaki08rps}. Due to these reasons, it is not immediately clear that
the picture of noise-sustained oscillations linked to patterns via spontaneous
synchronizations would be captured by such a treatment. In any case, this
kind of analysis offers an intriguing issue for future studies.

\section{Conclusion}

A model of two-species dynamics defined and treated in the language of 
hosts and parasitoids but equally applicable to prey--predator systems 
as well has been studied. In addition to extinct states, the system shows 
two kinds of coexistence, depending on the parameters. These are a patterned 
state with oscillating population densities and a non-patterned state with 
populations homogeneously distributed in space, and densities that fluctuate 
around their respective averages. This contribution contains two closely tied 
lines of work. One characterizes the patterns via coarse-graining the habitat
space to domains in which a given species or empty space is dominant. The other
concentrates on the dynamics of the system.

The coarse-grained domains allow for a definition of domain walls as the lines 
separating the domains, and vortices as their corner points. From these, we 
compute several quantities, such as the instantaneous vortex velocities, their 
lifetime, the lengths of the domain walls, and ones describing the geometry of 
the domains in the vicinity of the vortices. Invariably, these quantities 
reveal the division of the parameter space into two kinds of coexistence. We 
have argued that these two regions are separated by a continuous cross-over 
rather than a conventional phase transition. We have also reasoned why the 
vortex number process is not comparable to that in the CGLE 
\cite{aranson02,gil90,wang04}. The main argument is based on the fact that the 
process is not memoryless.

The dynamics has been studied using Poincar\'e maps \cite{zaslavsky}.
We have found out that as a stochastic dynamical 
system, the present one is two-dimensional. In other words, the dynamics is 
adequately described in terms of the two population densities and the unit-time
advancing mapping, the Poincar\'e map, of the two. No further dynamical 
variables nor history-dependence are needed. In addition, for large systems the
Poincar\'e maps are linear leading to an easy characterization of the behavior 
using eigenvalue analysis. In both the patterned and the non-patterned case, 
this reveals a stable fixed point with oscillatory transients. However, between
the cases there is a huge difference in the level of separation between the 
associated oscillation and decay time scales. This also leads to a direct 
connection between the patterning and the dynamics, since the ratio of the two 
timescales has been shown to be closely related to the average domain wall 
lengths. The amplitude fluctuations have also been given a spatial 
interpretation as an irregular sequence of spontaneous synchronization and 
desynchronization events of coupled oscillators \cite{acebron05} whose role in 
this setting is played by mesoscopic subsystems.

We have also compared the average densities in the system to those 
predicted by a mean-field theory. We have introduced corrections to the
MF equations on two levels based on series expansions and simulated data. 
In both zeroth- and first-order settings, the cross-over between the 
patterned and the non-patterned cases is visible and crucial. 

The results have been compared to earlier literature. There are numerous 
results on related systems where oscillations have been recovered. In some of 
these, the oscillations might in fact be noise-sustained in character. 
Interesting candidates to study this possibility include 
\cite{satulovsky94,tome07,arashiro08,antal01a,antal01b}. In all of these, the
machinery for pattern characterization introduced here could be applied to 
produce further insight. 

We have applied the analysis briefly to the empirical metapopulation landscape
of a butterfly and its parasitoids wasp \cite{nouhuys02,nouhuys04}. Future 
prospects include studying these issues further
- we have merely highlighted the applicability of the results presented here. 
Studying such systems could, for instance, let one approach the question of how
to tell the difference between patterns caused by inhomogeneities in the 
landscape and those that are spontaneously formed even on homogeneous 
substrates. Such extensions are naturally motivated also from the ecologists' 
point of view since metapopulation ecology often involves such landscapes 
\cite{hanski98}.

\subsection*{Acknowledgments} 

Ilkka Hanski and Otso Ovaskainen
(Univ.~Helsinki, Metapopulation Research Group)
are thanked for stimulating discussions
and Lasse Laurson for assistance. This work was supported by the
Academy of Finland through the Center of Excellence program (M.A.
and M.P.) and Deutsche Forschungsgemeinschaft via SFB 611 (M.R.).
The authors thank the Lorentz center (Leiden, Netherlands) for kind
hospitality.

\end{document}